\documentstyle[graphicx,amsmath,amssymb]{prepar}

\newcommand{\Teff}{$\rm{T_{eff}}$}
\newcommand{\FeH}{$[\rm Fe/H]$}
\newcommand{\alFe}{$[\alpha/\rm Fe]$}
\newcommand{\FeHsun}{$[\rm Fe/H]_\odot$}
\newcommand{\logg}{${\rm \log}~g$}
\newcommand{\g}{$g$}
\newcommand{\Y}{$Y$}
\newcommand{\Z}{$Z$}
\newcommand{\Mbol}{$\rm {M_{bol}}$}
\newcommand{\Fbol}{$\rm F_{\rm bol}$}

\newcommand{\sigMbol}{$\sigma_{\rm{M_{bol}}}$}
\newcommand{\sigpi}{$\sigma_\pi/\pi$~}
\newcommand{\sigFeH}{$\sigma_{[\rm Fe/H]}$~}
\newcommand{\aMLT}{${\alpha_{\rm MLT}}$~}
\newcommand{\aov}{${\alpha_{\rm ov}}$~}
\newcommand{\alf}{$\alpha$}
\newcommand{\Msun}{{$\rm M_{\odot}$}}
\newcommand{\HST}{{\sl HST}~}
\newcommand{\Mv}{{$\rm M_{\rm V}$}}
\newcommand{\dydz}{{$\Delta Y/\Delta Z$}~}
\newcommand{\dydzsun}{${(\Delta Y/\Delta Z)}_\odot$}
\newcommand{\Yp}{{${Y_{\rm p}}$}}

\begin{document}

\title{STELLAR STRUCTURE AND EVOLUTION: DEDUCTIONS FROM HIPPARCOS}

\markboth{Yveline Lebreton}{HIPPARCOS \& STELLAR ASTROPHYSICS}

\author{Yveline Lebreton
\affiliation{Observatoire de Paris,
DASGAL-UMR CNRS 8633, Place J. Janssen, 92195 Meudon, France\\
e-mail: Yveline.Lebreton@obspm.fr\\
--------------------\\
}}


\begin{abstract}

During the last decade, the understanding of fine features of
the structure and evolution of stars has become possible as a
result of
enormous progress made in the acquisition of high-quality
observational and experimental data and of new developments and
refinements in the theoretical description of stellar plasmas.
The confrontation of high-quality observations with
sophisticated stellar models has allowed many
aspects of the theory to be validated, and several characteristics of
stars relevant to Galactic evolution and cosmology to be inferred.
This paper is a review of the results of recent studies
undertaken in the context of the Hipparcos mission, taking
benefit of the high-quality astrometric data it has provided.
Successes are discussed, as well as the problems that have arisen
and suggestions proposed to solve them. Future observational and
theoretical developments expected and required in the field
are also presented.

\end{abstract}

\maketitle

\section{INTRODUCTION}
\label{intro}

Stars are the main constituents of the observable Universe.
The temperatures and pressures deep in their interiors are out of
reach for the observer, while the
description of stellar plasmas requires extensive knowledge in
various domains of modern physics such as nuclear and particle physics,
atomic and molecular physics, thermo- and hydrodynamics, physics
of the radiation and of its interaction with matter, and radiative transfer.
The development of numerical codes to calculate models of
stellar structure and evolution began more than forty years ago
with the pioneering works of Schwarzschild (1958) and Henyey et al (1959).
These programs have allowed at least the qualitative study and
understanding of numerous physical processes that intervene
during the various stages of stellar formation and evolution.

During the last two decades, observational data of increasingly
high accuracy have been obtained as a result of 1) the coming of
modern ground-based or space telescopes equipped with
high-quality instrumentation and with detectors giving access to
almost any possible range of wavelengths and 2) 
the elaboration of various sophisticated techniques of data reduction.
Ground-based astrometry has progressed, while space
astrometry was initiated with Hipparcos.
In the meantime, CCD detectors on large
telescopes opened the era of high-resolution, high signal-to-noise ratio
spectroscopy while multi-color filters were designed for photometry.
New fields have appeared or are under development, such as helio- and
asteroseismology or interferometry. On the other hand, stellar
models have been enriched by a continuously improved
physical description of the stellar plasma, while the use of
increasingly powerful computers has led to a gain in 
numerical accuracy.

The confrontation of models with observations allows testing and
even validation of the input physics of the models if numerous
observations of high quality are available. Fundamental returns are
expected in many domains that make use of quantitative results of
the stellar evolution theory such as stellar, Galactic, and
extragalactic astrophysics as well as cosmology. Because of their
positions, movements, or interactions with the interstellar medium
stars are actors and tracers of the dynamical and chemical
evolution of the Galaxy. Astrophysicists aim to determine their
ages and chemical compositions precisely. For example, the
firm determination of the ages of the oldest stars,
halo stars or members of globular clusters, is a long-standing
objective because it is one of the strongest constraints
for cosmology.

Although great progress has been made, a number of observations cannot
be reproduced by stellar models, which raises many questions regarding
both the observations and the models.
In the last few years, two scientific meetings have been explicitly devoted
to unsolved problems in stellar structure and evolution (Noels et
al 1995, Livio 2000). A major point of concern is that of transport processes
at work in stellar interiors (transport of the
chemical elements, angular momentum or magnetic fields by
microscopic diffusion and/or macroscopic motions).
Observations show that transport processes are indeed playing a
role in stellar evolution
but many aspects remain unclear (sometimes even unknown) and need
to be better characterized. Another crucial point concerns the
atmospheres, which link the stellar interior model to the interstellar medium and
are the intermediate agent between the star and the observer.
Uncertainties and inconsistencies in 
atmospheric descriptions generate errors in the analysis of
observational data and in model predictions.

This paper is the third of the series in ARAA dedicated to the
results of the Hipparcos mission; Kovalevsky
(1998) presented the products of the mission and the
very first astrophysical results obtained immediately after the
release of the data, while Reid (1999) reviewed 
the implications of the Hipparcos parallaxes
for the location of the main sequence (MS) in the Hertzsprung-Russell (H-R)
diagram, the luminosity calibration of primary
distance indicators, and the Galactic distance scale. Also,
van Leeuwen (1997) presented the results of the mission, and 
Baglin (1999) and Lebreton (2000) discussed the impact
of Hipparcos data on stellar structure and evolution.

Hipparcos has provided opportunities to study rather large
and homogeneous samples of stars sharing similar properties,
for instance, in terms of their space location or chemical composition.
I review studies based on Hipparcos observations
which ({\it 1}) confirmed several elements of stellar internal
structure theory, ({\it 2}) revealed some problems related to the
development of stellar models, and ({\it 3}) yielded more precise
characteristics of individual stars and clusters.
In Sections 2 and 3, I discuss the recent observational (including Hipparcos)
and theoretical developments from which new studies could be undertaken.
In Section 4, I concentrate on the nearest stars,
observed with highest precision (A-K disk and halo single or binary field
stars, and members of open clusters). In Section 5, I review recent results on
variable stars, globular clusters and white dwarfs based on
Hipparcos data. The stars considered are mostly of low
or intermediate mass, and except for white dwarfs, the evolutionary
stages cover the main sequence and subgiant branch.
Throughout this paper, I emphasize that the smaller error
bars on distances that result from Hipparcos make
the uncertainties on the other fundamental stellar parameters
more evident; fluxes, effective temperatures, abundances,
gravities, masses and radii have to be improved correspondingly,
implying in many cases the need for progress in atmospheric description.

\section{NEW HIGH-ACCURACY OBSERVATIONAL MATERIAL}
\label{obs}

This section presents a brief review of Hipparcos results 
and complementary ground-based or space observations 
which, if combined, provide very
homogeneous and precise sets of data.

\subsection{Space Astrometry with Hipparcos}
\label{hipp}

The Hipparcos satellite designed by the European Space Agency was
launched in 1989. The mission ended in 1993 and was followed by 3
years of data reduction. The contents of the Hipparcos Catalogue
(Eur. Space Agency 1997) were described by Perryman et al
(1997a). The data were released
to the astrophysical community in June~1997. General
information on the mission is given 
in van Leeuwen's (1997) and Kovalevsky's (1998) review papers.

Stars of various masses, chemical compositions and evolutionary stages
located either in the Galactic disk or in the halo were observed;
this was done systematically to a V-magnitude that 
depends on the galactic latitude and spectral type of the star,
and more generally, with a limit of V$\sim$12.4 mag. 
The Hipparcos Catalogue lists
positions, proper motions, and trigonometric parallaxes of 117\,955
stars as well as the
intermediate astrometric data, from which the astrometric
solutions were derived; this allows 
alternative solutions for the astrometric parameters to be
reconstructed according to different hypotheses
(see van Leeuwen \& Evans 1998).

A total of 12\,195 double or multiple systems are resolved
(among which about 25 \% were previously classified as single
stars), and 8\,542 additional stars are suspected to be
non-single. Detailed information on multiple systems, as described
by Lindegren et al (1997), can be
found in The Double and Multiple System Annex of the Catalogue.

The median accuracy on positions and parallaxes ($\pi$) is
typically $\sim$1 milliarcsecond (1~mas), whereas precisions on
proper motions are about 1 mas per yr.
Precisions become much higher for bright stars and worsen
toward the ecliptic plane and for fainter stars.
The astrometric accuracy and formal precision of Hipparcos data have been
investigated by Arenou et al (1995) and Lindegren (1995), and discussed
by van Leeuwen (1999a): for the Catalogue as a whole, the
zero-point error on parallaxes is below 0.1 mas and
the formal errors are not underestimated by more than 10\%.
After Hipparcos about 5\,200 single stars and 450
double (or multiple) stars have parallaxes known with an
accuracy \sigpi better than 5\%, 20\,853 stars
have \sigpi lower than 10\% and 49\,333 stars
have \sigpi lower than 20\% (Mignard 1997).
Martin et al (1997, 1998) and Martin \& Mignard (1998) determined the
masses of 74 astrometric binaries with accuracies in the range 5-35\%.
Söderhjelm (1999) obtained masses and improved orbital elements
for 205 visual binaries from a combination of Hipparcos astrometry
and ground-based observations; among these, 12 (20) systems have mass-errors
below 5 (7.5)\%.

The Hipparcos Catalogue also includes detailed and homogeneous
photometric data for each star, obtained from an average
number of 110 observations per star. The broad-band Hipparcos (Hp)
magnitude corresponding to the specific passband of
the instrument spanning the wavelength interval $\sim$ 350-800
nanometers (see Figure 1 in van Leeuwen et al 1997), is provided with
a median precision of 0.0015 mag for Hp$<$9 mag. The Johnson V magnitude
derived from combined satellite and ground-based observations is
given with a typical accuracy of 0.01 mag. The star mapper Tycho
had passbands close to the Johnson B and V bands and provided
two-color $\rm B_{T}$ and $\rm V_{T}$ magnitudes
(accuracies are typically 0.014 mag and 0.012 mag for stars with $\rm V_{T}<9$).

Hipparcos provided a detailed variability classification
of stars (van Leeuwen et al 1997), resulting in 
11\,597 variable or possibly variable stars. Among these
2\,712 stars are periodic variables (970 new) including 273 Cepheids, 186
RR Lyrae, 108 $\delta$ Scuti or SX Phoenicis stars, and 917
eclipsing binaries.

Hipparcos was planned more than fifteen years ago, and while its
development proceeded, significant progress was made in the
derivation of ground-based parallaxes using CCD detectors.
Parallaxes with errors less than 1.4 mas have already been obtained for
a few tens of stars, and errors are expected to drop to $\pm$0.5
mas in the years to come (Harris et al 1997, Gatewood et al 1998).
In addition, the Hubble Space Telescope (\HST) Fine Guidance Sensor
observations can provide parallaxes down to V$\sim$15.8
with errors at the 1 mas level (Benedict et al 1994, 
Harrison et al 1999). However, the distances of a
rather small number of stars will be measured by \HST because of
the limited observing time available for astrometry.
The enormous advantage of Hipparcos
resides in the large number of stars it dealt
with, providing homogeneous trigonometric parallaxes
that are essentially absolute.

\subsection{Ground-Based Photometry and Spectroscopy}
\label{GB}

The fundamental stellar parameters (bolometric magnitude \Mbol,
effective temperature \Teff, surface gravity \g, and chemical
composition) can be determined from photometry and/or
from detailed spectroscopic analysis.
However, the determination largely relies on 
model atmospheres and sometimes uses results of interior models.
Direct masses and radii can be obtained for stars belonging
to binary or multiple systems. Interferometry combined with
distances yields stellar
diameters giving direct access to \Teff, but still for a very
limited number of rather bright stars which
then serve to calibrate other methods. The different
methods (and related uncertainties) used to determine the
fundamental stellar parameters mainly for A to K Galactic dwarfs and
subgiants are briefly discussed, and improvements brought by Hipparcos
are underlined.

\noindent BOLOMETRIC MAGNITUDES.\ \
Integration of UBVRIJHKL photometry gives access to the bolometric
flux on Earth \Fbol, at least for F-G-K stars where most energy
is emitted in those bands and
which are close enough not to be affected by interstellar absorption;
the (small) residual flux,
emitted outside the bands, is estimated from
model atmospheres. Recently, Alonso et al (1995) applied
the method to $\sim$100 F-K dwarfs and subdwarfs and obtained
bolometric fluxes accurate to about 2\% and, as a by-product,
empirical bolometric corrections for MS stars.

If \Fbol~ and distances are known, \Mbol~ can be derived
with no need for bolometric correction. The accuracy is then
$\sigma_{\rm M_{\rm bol}}=\log e\ {[{(2.5\frac{\sigma_{\rm F_{\rm bol}}}
{\rm F_{\rm bol}})^2} +
{(5 \frac{\sigma_\pi}{\pi} )^2}]}^\frac{1}{2}~$, meaning that if
$\frac{\sigma_{\rm F_{\rm bol}}}{\rm F_{\rm bol}}\sim$2\% then
$\sigma_{\rm M_{\rm bol}}$ is dominated by
the parallax error as soon as $\frac{\sigma_\pi}{\pi}>$1\%.
In other cases, when the distance is
known, \Mbol~ is obtained from any apparent magnitude $m$ and its
corresponding bolometric correction BC($m$), derived from empirical
calibrations or from model atmospheres. Up to now Hipparcos magnitudes
Hp have not been used extensively, despite their excellent accuracy
(0.0015~mag), because of remaining difficulties in calculating BC(Hp)
(Cayrel et al 1997a).

\noindent EFFECTIVE TEMPERATURES.\ \
The InfraRed Flux Method (IRFM; Blackwell et al 1990), applicable
to A-K stars proceeds in two steps.
First, the stellar angular diameter
$\phi$ is evaluated by comparing the IR flux observed on Earth
in a given band to the flux predicted by a model atmosphere
calculated with the observed gravity and abundances and an approximate \Teff~ (the IR flux
does not depend sensitively on \Teff).
Then \Teff~ is obtained from the total (integrated) flux \Fbol~ and
$\phi$. Iteration of the procedure yields a ``definite''
value of \Teff. Using IRFM, Alonso et al (1996a) derived
temperatures of 475 F0-K5 stars (\Teff~ in the range 4000-8000K) with
internal accuracies of $\sim$1.5\%. The zero-point of their \Teff-scale
is based on direct interferometric measures by Code et al
(1976), and the resulting systematic uncertainty is $\sim$1\%.
Accuracies of $\sim$1\% were obtained by Blackwell \& Linas-Gray (1998),
who applied IRFM to 420 A0-K3 stars, corrected for interstellar
extinction using Hipparcos parallaxes. Both sets of results 
compare well, with differences below 0.12$\pm$1.25\% for the 93 stars in common.

The surface brightness method (Barnes et al 1978) was applied
by Di Benedetto (1998) to obtain a (\Teff, V-K) calibration.
The calibration is based on 327 stars with high-precision K-magnitudes
from the Infrared Space Observatory (ISO), Hipparcos V-magnitudes and
parallaxes (the latter to correct for interstellar extinction), and
bolometric fluxes from Blackwell \& Linas-Gray (1998).
First, the visual surface brightness
\hbox{$S_V=\rm V+5\log\phi$} is calibrated as a function of (V-K) using 
stars with precise $\phi$ from interferometry. Then for any 
star $S_V$ is obtained from (V-K), $\phi$ from $S_V$ and V,
yielding in turn \Teff~ from \Fbol~ and $\phi$.
From the resulting (\Teff, V-K) calibration, Di Benedetto
derived \Teff~values of 537 ISO A-K dwarfs and giants with $\pm$1\%
accuracy. The method produces results in good agreement with
those of IRFM and is less dependent on atmosphere
models.

Multiparametric empirical calibrations of \Teff~
as a function of the color indices and eventually of
metallicity \FeH~ (logarithm of the number abundances of Fe to H
relative to the solar value) and gravity can be derived from the 
empirical determinations of the effective temperatures of the rather nearby
stars. In turn, the effective temperature of any star lying in the
(rather narrow) region of the H-R diagram covered by a given
calibration can easily be derived (see for example Alonso et al 1996b).
Empirical calibrations also serve to validate purely
theoretical calibrations based on model atmospheres; these latter
have the advantage of covering the entire parameter space of the
H-R diagram (i.e. wide ranges of color indices, metallicities and
gravities; see Section~\ref{atmos} later in this article).

Spectroscopic determination of \Teff~ is based on the analysis of chosen
spectral lines that are sensitive to temperature; for
instance the Balmer lines for stars with \Teff~ in the interval 5000-8000~K.
Because of the present high quality of the stellar spectra,
precisions of $\pm$50-80~K on \Teff, that correspond to the adjustment of the
theoretical line profile to the observed one, are commonly found in the
literature (Cayrel de Strobel et al 1997b, Fuhrmann 1998).
This supposes that theoretical profiles are very
accurate, and therefore neglects the model atmosphere uncertainties.

Popper (1998) used detached eclipsing binaries with rather
good Hipparcos parallaxes, accurate radii, and measured V-flux to
calibrate the radiative flux as a function of (B-V); he found good
agreement with similar calibrations based on interferometric angular diameters.
From the same data, Ribas et al (1998) derived effective temperatures
(this required bolometric corrections) and found them to be in reasonable
agreement (although systematically smaller by 2-3\%)
with \Teff~ derived from photometric calibrations.
However the stars are rather distant, which implies rather significant
internal errors on \Mbol~ and \Teff~ (a parallax error of 10\%
is alone responsible for a \Teff-error of 5\%). In Ribas et al's sample,
only a few systems have \sigpi$<$10\%, and because errors on radius,
magnitudes, and BC also intervene, only 5 systems
have \Teff~ determined to better than 3\%.

\noindent SURFACE GRAVITIES. \ \
If \Teff~ and \Mbol~ are known, the radius of the star may be
derived from the Stefan-Boltzmann law and the mass estimated from a grid of
stellar evolutionary models, yielding in turn the value of \g.
This method has been applied to a hundred metal-poor subdwarfs
and subgiants with accurate distances from Hipparcos
(Nissen et al 1997, Fuhrmann 1998, Clementini et al 1999).
Nissen et al showed that among the various sources of errors,
the error on distance still dominates, but pointed out that if
the distance error is lower than 20\% then the error on \logg~ may be
lower than $\pm$~0.20 dex.

On the other hand, \g\ can be determined from spectroscopy.
Different gravities produce different atmospheric pressures ,
modifying the profiles of some spectral lines.
Two methods have been widely used to estimate \g.
The first method is based on the analysis of the equation of
ionization equilibrium of abundant species, iron,
for instance. The iron abundance is determined from FeI lines
that are not sensitive to gravity, and then \g\ is adjusted so
that the analysis of FeII lines, which are sensitive to gravity,
leads to the same value of the iron abundance. The
accuracy in \logg~ is in the range $\pm$0.1-0.2 dex (Axer et al
1994). The second method relies on the analysis of the wings of strong lines
broadened by collisional damping, such as Ca I (Cayrel et al 1996) or
the Mg Ib triplet (Fuhrmann et al 1997), leading to uncertainties smaller than 0.15
dex. The two methods often produce quite different results, with
systematic differences of $\sim$0.2-0.4 dex, at least
when ionization equilibria are estimated from models in local
thermodynamical equilibrium (LTE).

Th\'evenin \& Idiart (1999) have studied the effects of
departures from LTE on the formation of FeI and FeII lines in
stellar atmospheres, and found that modifications of the ionization
equilibria resulted from the overionization of iron induced by significant UV fluxes.
The nice consequence is that the gravities they inferred from iron ionization
equilibrium for a sample of 136 stars
spanning a large range of metallicities become very close to gravities
derived either from pressure-broadened strong
lines or through Hipparcos parallaxes.

\noindent ABUNDANCES OF THE CHEMICAL ELEMENTS. \ \
The spectroscopic determination of abundances of chemical elements rests
on the comparison of the outputs of model atmospheres (synthetic spectra,
equivalent widths) with their counterpart in the observed
spectra. This requires a preliminary estimate of \Teff~ and \g.
If high-resolution spectra are used, the line widths are very precise
and the internal uncertainty in abundance determinations depends on
uncertainties in \g\ and \Teff, on the validity of the model atmosphere,
and on the oscillator strengths. Error bars in the range
$\pm~$0.05-0.15~dex are typical (Cayrel de Strobel et al 1997b, Fuhrmann 1998).
Also when different sets of \FeH~ determinations are compared,
the solar Fe/H ratio used as reference
must be considered; values differing
by $\sim$0.15 dex used to be found in the literature (Axer et
al 1994). This has resulted in long-standing difficulties in determining
the solar iron abundance from FeI or FeII lines, because of
uncertain atomic data. In a recent paper, Grevesse \& Sauval
(1999) reviewed the problem and opted for a ``low''
Fe-value, $\rm A_{Fe}=7.50\pm0.05$ ($\rm A_{Fe}=\log(n_{Fe}/n_H)+12$
is the logarithm of the number density ratio of Fe to H
particles), in perfect agreement with the meteoritic value.

Furthermore, if abundances are estimated from model atmospheres in
LTE, perturbations of statistical equilibrium by
the radiation field are neglected.  Th\'evenin \& Idiart (1999)
found that in metal-deficient dwarfs and subgiants, the iron overionization
resulting from reinforced UV flux modifies the line widths. They
obtained differential non-LTE/LTE abundance corrections
increasing from 0.0 dex at \FeH =0.0 to +0.3 dex at \FeH = -3.0.
These corrections are indeed supported by the agreement
between spectroscopic gravities and ``Hipparcos'' gravities
discussed previously.

{\sl Helium} lines do not form in the photosphere of low-temperature stars
which precludes a direct determination of helium abundance. The
calibration of the solar model in luminosity and radius at solar
age yields the initial helium content of the Sun (Christensen-Dalsgaard
1982), while oscillation frequencies give access to the present
value in the convection zone (Kosovichev et al 1992). In other stars, it is
common to use the well-known scaling relation $Y -Y_p
= Z \frac{\Delta Y}{\Delta Z}$, which supposes that the helium
abundance has grown with metallicity \Z~ from the primordial
value $Y_p$ to its stellar birth value \Y ~(\Y~ and \Z~ represent
abundances in mass fraction); \dydz is the enrichment factor.

{\sl \alf-element} abundances (O, Ne, Mg, Si, S, Ar, Ca, Ti) have now
been widely measured in metal-deficient stars.
Stars with \FeH$\lesssim$-0.5 dex generally exhibit an
\alf-element enhancement with respect to the Sun (\alFe)
quite independent of their metallicity
(Wheeler et al 1989, Mc William 1997). Recent determinations of \alFe~ in
99 dwarfs with \FeH$<$-0.5 from high-resolution
spectra by Clementini et al (1999) yield \alFe=+0.26$\pm$0.08 dex.

\section{RECENT THEORETICAL AND NUMERICAL PROGRESS}
\label{theory}

Recent developments in the physical description of low and
intermediate mass stars are briefly presented.

\subsection{Microscopic Physics}
\label{micphy}

The understanding of stellar structure benefited substantially 
from the complete re-examination of stellar opacities by two groups:
the Opacity Project (OP, see Seaton et al 1994) and the OPAL group
at Livermore (see Rogers \& Iglesias 1992). Both showed by
adopting different and independent approaches, that improved
atomic physics lead to opacities generally higher than the
previously almost ``universally'' used Los Alamos opacities
(Huebner et al 1977). The opacity enhancements reach factors of 2-3
in stellar envelopes with temperatures in the range $10^5-10^6$~K.
With these new opacities, ({\it 1}) a number of long-standing problems in 
stellar evolution have been solved or at least lessened and ({\it 2})
finer tests of stellar structure could be undertaken.
Since opacity is very sensitive to metallicity, any underlying
uncertainty on metallicity may be problematic.

Great efforts have also been invested in the derivation of low-temperature opacities,
including millions of molecular and atomic lines and grain absorption
that are fundamental for the calculation of the envelopes
and atmospheres of cool stars (Kurucz 1991, Alexander \& Ferguson 1994).

OP and OPAL opacities have been shown to be in reasonable agreement (Seaton et
al 1994, Iglesias \& Rogers 1996); and very good agreement between
OPAL and Alexander \& Ferguson's or Kurucz's
opacities is found in the domains where they overlap.
Although some uncertainties remain that are difficult to
quantify, the largest discrepancies between the various
sets of tables do not exceed 20\% and are generally well
understood (Iglesias \& Rogers), making opacities 
much more reliable today than they were ten years ago.

The re-calculation of opacities required appropriate
equations of state (EOS). The MH\&D
EOS (see Mihalas et al 1988) is part of the OP project,
while the OPAL EOS was developed at
Livermore (Rogers et al 1996). In the meantime,	another
EOS was designed to interpret the first observations of
very low-mass stars and brown dwarfs (Saumon \& Chabrier 1991).
OPAL and OP EOS are needed to satisfy the strong helioseismic
constraints (Christensen-Dalsgaard \& Däppen 1992).

\subsection{Atmospheres}
\label{atmos}

Atmospheres intervene at many levels in the analysis of
observations (Section~\ref{GB}). They also
provide external boundary conditions for the
calculation of stellar structure and necessary relations
to transform theoretical (\Mbol, \Teff) H-R diagrams to
color-magnitude (C-M) or color-color planes.
Models have improved during the last two decades, and
attention has been paid to the treatment of atomic and molecular
line blanketing.  The original programs MARCS of Gustafsson et al (1975) and
ATLAS by Kurucz (1979) evolved toward the most recent
ATLAS9 version, appropriate for O-K stars (Kurucz 1993) and NMARCS
for A-M stars (see Brett 1995 and Bessell et al 1998).
On the other hand; very low-mass stellar atmosphere models
were developed;
Carbon (1979) and Allard et al (1997) reviewed calculation details and 
remaining problems (such as incomplete opacity data, poor
treatment of convection, neglect of non-LTE effects or
assumption of plane-parallel geometry).

\noindent COLOR-MAGNITUDE TRANSFORMATIONS.\  \
Different sets of transformations (empirical or theoretical)
were used to analyze the ``Hipparcos'' stars.
Empirical transformations have been discussed in
Section~\ref{GB}. The most recent theoretical
transformations are compiled by Bessell et al (1998),
who used synthetic spectra derived from ATLAS9
and NMARCS to produce broad-band colors and bolometric
corrections for a very wide range of \Teff, \g\ and [Fe/H] values.
These authors found fairly good agreement with empirical relations
except for the coolest stars (M dwarfs, K-M giants).

\noindent INTERIOR/ATMOSPHERE INTERFACE.\ \
The external boundary conditions for interior models are
commonly obtained from $\rm T(\tau)$-laws ($\tau$ is the optical
depth) derived either from theory or full atmosphere calculation.
This method is suitable for low- and intermediate-mass stars
(it is not valid for masses below $\sim$0.6~\Msun, Chabrier \& Baraffe 1997).
Morel et al (1994) and Bernkopf (1998) focused on the solar case
where seismic constraints require a careful handling of
external boundary conditions. Morel et al pointed out that homogeneous
physics should be used in interior and atmosphere
(opacities, EOS, treatment of convection) and
showed that the boundary level must be set deep enough, in zones where the
diffusion approximation is valid. Bernkopf
discussed some difficulties in reproducing Balmer lines related to the
convection treatment.

\subsection{Transport Processes}
\label{transp}

\noindent CONVECTION.\ \ 3-D numerical simulations at current numerical
resolution are able to reproduce most observational features
of solar convection such as images, spectra, and helioseismic properties
(Stein \& Nordlund 1998). However, the ``connection'' with a stellar
evolution code is not easy, and stellar models still mostly
rely on 1-D phenomenological descriptions such as the
mixing-length theory of convection (MLT, B\"ohm-Vitense 1958). The mixing-length parameter
\aMLT (ratio of the mixing-length to the pressure scale height) 
is calibrated so that the solar model yields the observed
solar radius at the present solar age. The question of the variations of
\aMLT in stars of various masses, metallicities, and evolutionary
stages remains a matter of debate (Section~\ref{best}).
As pointed out by Abbett et al (1997), the MLT can reproduce the
correct entropy jump across the superadiabatic layer near the
stellar surface, but fails to describe the detailed depth
structure and dynamics of convection zones. Abbett et al found that
the solar entropy jump obtained in 3-D simulations corresponds to
predictions of the MLT for \aMLT$\thickapprox$1.5.
Ludwig et al (1999) calibrated \aMLT from 2-D simulations of compressible convection
in solar-type stars for a broad range of \Teff~ and \g-values.
The solar \aMLT inferred from 3-D and 2-D simulations is close to what is
obtained in solar model calibration. The \aMLT
dependence with \Teff~ and \g\ of Ludwig et al can be used to
constrain the range of acceptable variations of \aMLT in
stellar models (see Section~\ref{best}).

\noindent OVERSHOOTING.\  \ Penetration of convection and mixing beyond
the classical Schwarzschild convection cores (overshooting
process) modifies the standard evolution model of stars of masses
M$\gtrsim$1.2 \Msun, in particular the lifetimes (see for instance 
Maeder \& Mermilliod 1981, Bressan et al 1981).
The extent of overshooting was estimated for the first time from the comparison of
observed and theoretical
MS widths of open clusters (Maeder \& Mermilliod 1981),
which yields an overshooting parameter \aov$\sim$0.2
(ratio of overshooting distance to pressure scale height).
As discussed in detail by Roxburgh (1997), \aov is still poorly
constrained despite significant efforts made to establish the dependence of
overshooting with mass, evolutionary stage, or chemical
composition (see Section~\ref{best}). Andersen (1991) first pointed out that
the simultaneous calibration of well-known binaries
(masses and radii at 1-2\%) may provide improved
constraints for \aov. A modeling of the sample
of the best-known binaries indicates a trend for
\aov to increase with mass and suggests a decrease of \aov with decreasing metallicity
(Ribas 1999), although a larger sample would be desirable to
confirm those trends. Further advances are expected from
asteroseismology (Brown et al 1994, Lebreton et al 1995).

\noindent DIFFUSION OF CHEMICAL ELEMENTS.\ \  Various mixing processes
may occur in stellar radiative zones (see Pinsonneault 1997).
In low-mass stars, microscopic diffusion due to gravitational
settling carries helium and heavy elements down to the center and modifies
the evolutionary course as well as the surface abundances.
It has been proved that microscopic diffusion can explain the low
helium abundance of the solar convective zone derived from
seismology (Christensen-Dalsgaard et al 1993).
On the other hand, turbulent mixing  (resulting, for instance, from hydrodynamical
instabilities related to rotation, see Zahn 1992) probably inhibits
microscopic diffusion. Richard et al (1996) did not find any
conflict between solar models including rotation-induced mixing (to
account for Li and Be depletion at the surface) and
microscopic diffusion (to account for helioseismic data).
More constraints are required to clearly identify (and quantify the effects of)
the various candidate mixing processes; this will be
illustrated in the following sections.

\section{STUDIES OF THE BEST-KNOWN OBJECTS}
\label{best}

Stellar model results depend on a number of free input parameters. Some
are observational data (mass, chemical composition and age, the
latter for the Sun only), whereas others enter phenomenogical descriptions of
poorly-known physical processes (mixing-length parameter
for convection, overshooting, etc).
The model outputs have to be compared with the best
available observational data:
luminosity, \Teff~ or radius, oscillation frequencies, etc.
Numerous and precise observational constraints 
allow assessment of the input physics or give
more precise values of the free parameters.
They may reveal the necessity to include processes previously
neglected and in the best cases to characterize them.

The model validation rests on (1) the nearest objects with the
most accurate observations, (2) special objects with
additional information such as stars belonging to
binary systems, members of stellar clusters or stars with seismic
data, and (3) large samples of objects giving access to
statistical studies. 

\subsection{Stars in Binary Systems}
\label{bin}

Masses are available for a number of stars belonging to
binary systems, allowing their ``calibration''
under the reasonable assumption that the stars have the same age and
were born with the same chemical composition (Andersen 1991, Noels et al 1991).
A solution is sought which reproduces the
observed positions in the H-R diagram of both stars.
Andersen (1991) claimed that the only systems able to really constrain
the internal structure theory are those with errors lower than
2\% in mass, 1\% in radius, 2\% in \Teff~ and 25\%
in metallicity.

However, additional observations may sometimes
cast doubts on an observed quantity previously
determined with good internal accuracy. This occurred recently for
the masses of stars in the nearest visual binary
system \alf~Centauri. The system has been widely
modeled in the past (Noels et al 1991, Edmonds et al 1992, Lydon
et al 1993, Fernandes \& Neuforge 1995) with the objective of
getting (among others) constraints on the
mixing-length parameter. At that time, the astrometric masses
were used (internal error of 1\%) but the \FeH-value was
controversial, leading to various possibilities for \aMLT-values.
Today the situation is still confused: metallicity is better 
assessed, but new radial velocity measurements yield masses
higher than those derived from astrometry (by 6-7\%,
Pourbaix et al 1999). The higher masses imply a reduction in age by a
factor of 2 and slightly different \aMLT-values for the two stars.
However, the orbital parallax corresponding to the high-mass
``option'' is smaller than and outside the error bars of both ground-based
and Hipparcos parallax $\pi_{\rm Hipp}$. Pourbaix et al noted the
lack of reliability of $\pi_{\rm Hipp}$ given in the Hipparcos Catalogue,
but since then it has been re-determined from intermediate data
by Söderhjelm (1999) and is now close to (and in agreement with)
the ground-based parallax. More accurate radial velocity
measurements are therefore needed to assess the high-mass solution.

Possible variations of \aMLT have been investigated through the
simultaneous modeling of selected nearby visual binary systems
(Fernandes et al 1998, Pourbaix et al 1999, Morel et al 2000).
Small variations of \aMLT (not greater than $\approx$0.2)
in the two components of \alf Cen
(Pourbaix et al) and $\iota$ Peg (Morel et al) have been suggested.
Fernandes et al, who calibrated 4 systems and the Sun with the same
program and input physics, found that \aMLT is almost constant
for \FeH~ in the range \FeHsun$\pm$0.3 dex and masses between
0.6 and 1.3 \Msun.
In this mass range the sensitivity of models to \aMLT increases
with mass (due to the increase with mass of the entropy jump
across the superadiabatic layer) which makes the MS slope vary
with \aMLT. Also, I estimate from my models that a change of 
\aMLT of $\pm0.15$ around 1~\Msun~ translates into a \Teff-change
of $\sim$40-55~K depending on the metallicity. 
On the other hand, with the solar-\aMLT value
the MS slope of field stars and Hyades stars is well fitted
(Section~\ref{OC}). It is therefore reasonable to adopt the
solar-\aMLT value to model {\sl solar-type stars}. For other stars, the situation
is less clear. The calibration of \aMLT depends on the external
boundary condition applied to the model, itself sensitive to the
low-temperature opacities, and on the color transformation used
for the comparison with observations. Chieffi et al (1995) examined
the MS and red giant branch (RGB) in metal-deficient clusters and
suggested a constancy of \aMLT from MS to RGB and a decrease
with decreasing \Z. They found variations of \aMLT with \Z~ of $\approx$
0.2-0.4, but these are difficult to assess considering uncertainties
in the observed and theoretical cluster sequences.
On the other hand, calibration of \aMLT with 2-D simulations of convection gives
complex results (Freytag \& Salaris 1999; Freytag et al 1999). In particular,
(1) for solar metallicity, \aMLT is found to decrease when \Teff~ increases above solar
\Teff, and to increase slightly when stars move toward the RGB (by $\approx$
0.10-0.15) and; (2) \aMLT does not vary importantly when metallicity decreases at solar
\Teff.
More work is needed to go into finer details, and other
calibrators of \aMLT are required, such as binary stars in the appropriate
range of mass and with various chemical compositions.

The modeling of a moderately large
sample of binaries might give information on the variation of
helium \Y~ and age with metallicity \Z, of great interest
for Galactic evolution studies. The combined results for six binary
systems and the Sun with the same program by Fernandes et al
(1998) and Morel et al (2000) show a general trend for \Y~ to increase with
\Z: \Y~ increases from 0.25 to 0.30 ($\pm$0.02) when \Z~ increases
from 0.007 to 0.03 ($\pm$0.002). However, the Hyades appear to
depart from this tendency (see Section~\ref{OC}).

The sample of binaries with sufficiently accurate temperatures 
and abundances is still too meager to allow full characterization of
physical processes. Additional data are needed such as
observations of binaries in clusters (see Section~\ref{OC}) or
asteroseismological measurements.

\subsection{The Nearest Disk and Halo Stars}
\label{nearby}

\noindent FINE STRUCTURE OF THE H-R DIAGRAM.\ \
Highly accurate distances for a rather
large number of stars in the solar neighborhood were provided by Hipparcos.
This allowed the first studies of the fine structure
of the H-R diagram and related metallicity effects to be undertaken.

\begin{figure}
\resizebox{\hsize}{!}{\includegraphics{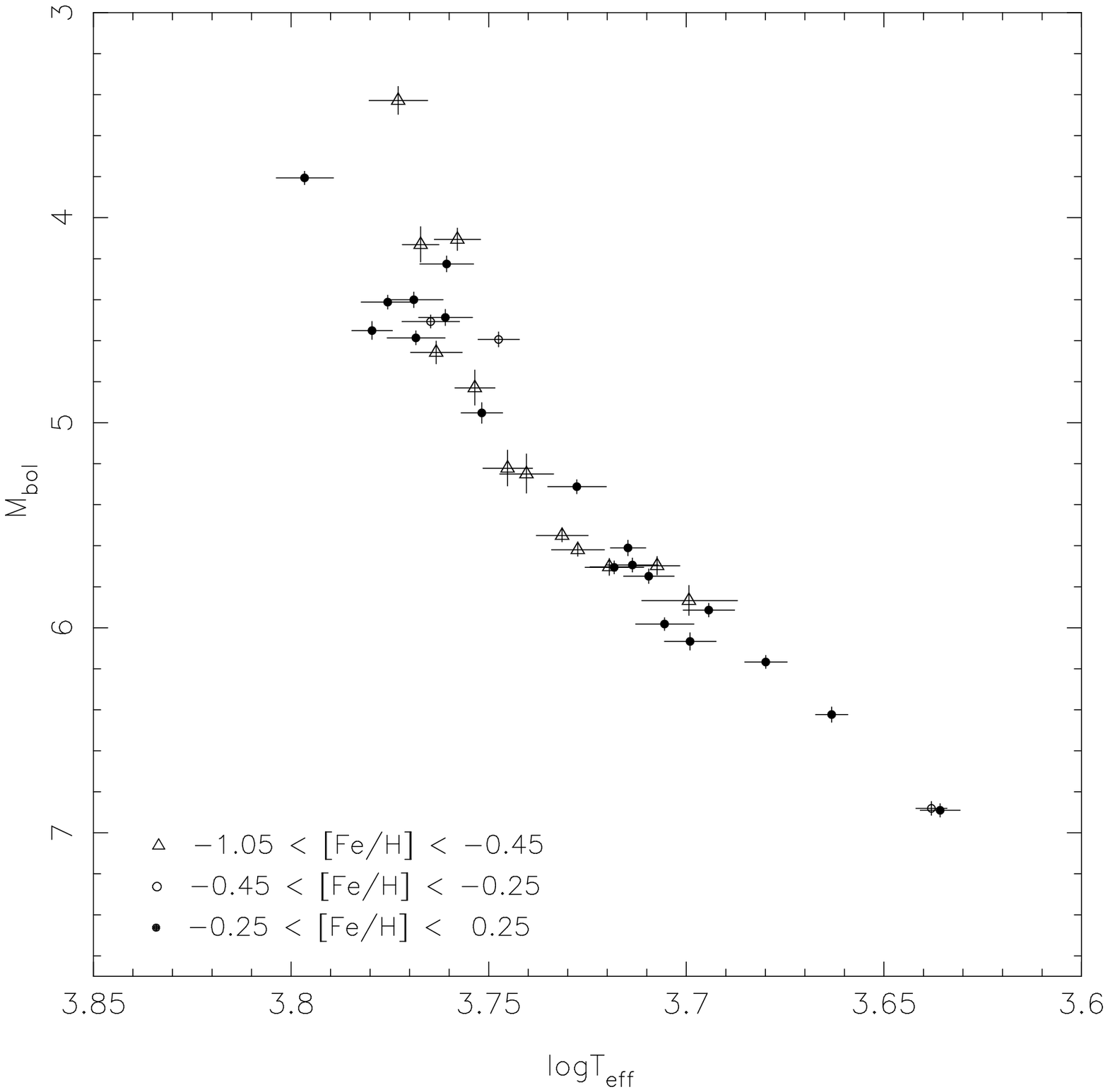}}

\caption{Hipparcos H-R  diagram of the 34 best-known nearby
stars. The parallax accuracies \sigpi are in the range
0.003-0.041.
Bolometric fluxes and effective temperatures are available from Alonso
et al's (1995, 1996a) works ($\frac{\sigma_{\rm F_{bol}}}{\rm F_{bol}}\sim 2\%$ and
$\frac{\sigma_{\rm T_{eff}}}{\rm T_{eff}}\sim 1.5\%$, see Section~\ref{GB}).
Resulting \sigMbol~ are in the range 0.031-0.095 mag (from Lebreton et al
1999).}
\label{LebrF1}
\end{figure}

Among an ensemble of ``Hipparcos'' F-G-K stars closer than 25 pc,
with error on parallax lower than 5\%, Lebreton et al (1997b)
selected stars with \FeH~ in the range $[-1.0,+0.3]$ from
detailed spectroscopic analysis (\sigFeH$\simeq$0.10\ dex, Cayrel de
Strobel et al 1997b), \Fbol~ and \Teff~ from Alonso et al (1995,
1996a) with $\frac{\sigma_{\rm F_{bol}}}{\rm F_{bol}}\sim 2\%$ and
$\frac{\sigma_{\rm T_{eff}}}{\rm T_{eff}}\sim 1.5\%$ (see Section~\ref{GB})
and not suspected to be unresolved binaries. 
Figure~\ref{LebrF1} presents the H-R diagram of the
34 selected stars: the error bars are the smallest obtained for stars
in the solar neighborhood (\sigMbol are in the range 0.031-0.095 with an average value
$\langle$\sigMbol$\rangle\simeq$0.045\ mag).
The sample is compared with theoretical isochrones
derived from standard stellar models in Figure~\ref{LebrF2}.
Models cover the entire \FeH-range. They account for an \alf-element enhancement
\alFe=+0.4 dex for \FeH$\leq$-0.5 and, for non-solar \FeH, 
have a solar-scaled helium content (\Y=\Yp+\Z \dydzsun).
The splitting of the sample into a solar metallicity sample and a
moderately metal-deficient one (Figure~\ref{LebrF2}a and b) shows that:
\begin{enumerate}
\item The slope of the MS is well reproduced with the solar \aMLT,
\item Stars of solar metallicity and close to it occupy the theoretical
band corresponding to their (LTE) metallicity range, while for moderately
metal deficient stars there is a poor fit.
\end{enumerate}

In general, stars have a tendency to lie on a theoretical
isochrone corresponding to a higher metallicity than
the spectroscopic (LTE) value.
This trend was already noticed by Axer et al (1995) but it is now
even more apparent because of the high accuracy of the data. Helium content
well below the primordial helium value would be required to resolve the conflict.

This is exemplified by the star $\mu$
Cas~A, the A-component of a well-known, moderately metal-deficient
binary system that has a well-determined mass (error in mass of 8 per cent).
The standard model (Figure~\ref{LebrF3}) is more than 200~K hotter
than the observed point and is unable to reproduce the observed \Teff~ even if
(reasonable) error bars are considered (Lebreton 2000).
On the other hand, the mass-luminosity
properties of the star are well reproduced if the helium
abundance is chosen to be close
to the primordial value, although the error bar in mass is
somewhat too large to provide strong constraints.

Several reasons can be invoked to explain the poor fit at low
metallicities:

\begin{enumerate}
\item {\sl Erroneous temperature-scale.} 3-D model atmospheres could
still change the \Teff-scale as a function of metallicity
(Gustafsson 1998), but with the presently (1-D) available models
it seems difficult to increase Alonso et al's (1996a)
\Teff~ by as much as 200-300~K. As noted by Nissen (1998), this
scale is already higher than other photometric
scales, by as much as 100~K. Also, Lebreton et al (1999)
verified that spectroscopic effective temperatures lead to a similar misfit.
\item {\sl Erroneous metallicities.} As discussed in Section~\ref{GB},
the \FeH-values inferred from model atmosphere analysis should
be corrected for non-LTE effects.
According to Th\'evenin \& Idiart (1999) no correction is
expected at solar metallicity, whereas for moderately metal-deficient
stars the correction amounts to $\sim$0.15\ dex.
\item {\sl Inappropriate interior models.} In low-mass stars,
microscopic diffusion by gravitational settling
can make helium and heavy elements sink toward the center, changing
surface abundances as well as inner abundance profiles.
In metal-deficient stars this process may be very efficient for
three reasons: (1) densities at the bottom of the convection zone
decrease with metallicity, which favors settling; (2) the thickness of
the convection zones decreases with metallicity, making the
reservoir easier to empty; and (3) metal-deficient generally means
older, which implies more time available for diffusion.
\end{enumerate}

The two latter reasons are attractive because they qualitatively
predict an increasing deviation from the standard case
when metallicity decreases. As shown in Figure~\ref{LebrF3},
a combination of microscopic diffusion effects with
non-LTE Fe/H corrections could remove the discrepancy
noted for $\mu$Cas A: an increase of \FeH~ by
0.15 dex produces a rightward shift of 80~K of the standard isochrone,
representing about one third of the discrepancy. Additionally, adding
microscopic diffusion effects, according to recent calculations by
Morel \& Baglin (1999), provides a match to the observed positions.
Moreover, the general agreement for solar metallicity stars
(Figure~\ref{LebrF2}a) should remain: (1) at solar metallicities non-LTE
corrections are found to be negligible, and (2) at ages of $\sim$5
Gyr chosen as a mean age for those (expectedly) younger stars, diffusion effects
are estimated to be smaller than the error bars on \Teff~ (Lebreton et al 1999).

To conclude on this point, the high-level
accuracy reached for a few tens of stars in the solar neighborhood
definitely reveals imperfections in interior and atmosphere models.
It  casts doubts on abundances derived from model atmospheres in
LTE, and favors models that include microscopic diffusion of helium and
heavy elements toward the interior over standard models.
Also, diffusion makes the surface \FeH-ratio decrease by $\sim$0.10 dex in 10 Gyr in a
star like $\mu$ Cas (Morel \& Baglin 1999), which is rather small
and hidden in the observational error bars. In very old,
very deficient stars, the \FeH-decrease is expected to be larger (Salaris et al 2000),
which makes
the relation between observed and initial abundances difficult to establish.
In the future, progress will come from the study of enlarged
samples reaching the same accuracies and of the
acquisition of additional parameters to constrain the models.
The knowledge of masses for several binaries in a narrow mass
range but large metallicity range would help to
constrain the helium abundances, while access to seismological
data for at least one or two stars would help to better characterize
mixing processes.

\begin{figure}
\resizebox{\hsize}{!}{\includegraphics{LebrF2.eps}}

\caption{The sample of Figure~\ref{LebrF1} 
split into two metallicity domains. Figure~\ref{LebrF2}a shows
stars with \FeH~ close to solar (\FeH~$\in$[-0.45, +0.25]).
Figure~\ref{LebrF2}b shows 
moderately metal-deficient stars (\FeH~$\in$ [-1.05,-0.45]).
Theoretical isochrones are overlaid on the observational data. 
Figure~\ref{LebrF2}a: the lower isochrone (10 Gyr) is for \FeH=-0.5,
\Y=0.256 and \alFe=0.4;
the upper isochrone (8 Gyr) is for \FeH=+0.3, \Y=0.32 and
\alFe=0.0; the dashed line is a solar ZAMS (\aMLT=1.65,
$Y_\odot$=0.266 and $Z_\odot$=0.0175). The brightest star is the young
star $\gamma$ Lep. Figure~\ref{LebrF2}b: 2 isochrones (10 Gyr)
with \alFe=0.4, the lower is for \FeH=-1.0, \Y=0.236 and the upper for
\FeH=-0.5, \Y=0.256. All stars but
one are sitting above the region defined by the isochrones.
\dydzsun=2.2 is obtained with Balbes et al's (1993)
primordial helium \Yp=0.227 (from Lebreton et al 1999).}

\label{LebrF2}
\end{figure}
\begin{figure}
\resizebox{\hsize}{!}{\includegraphics{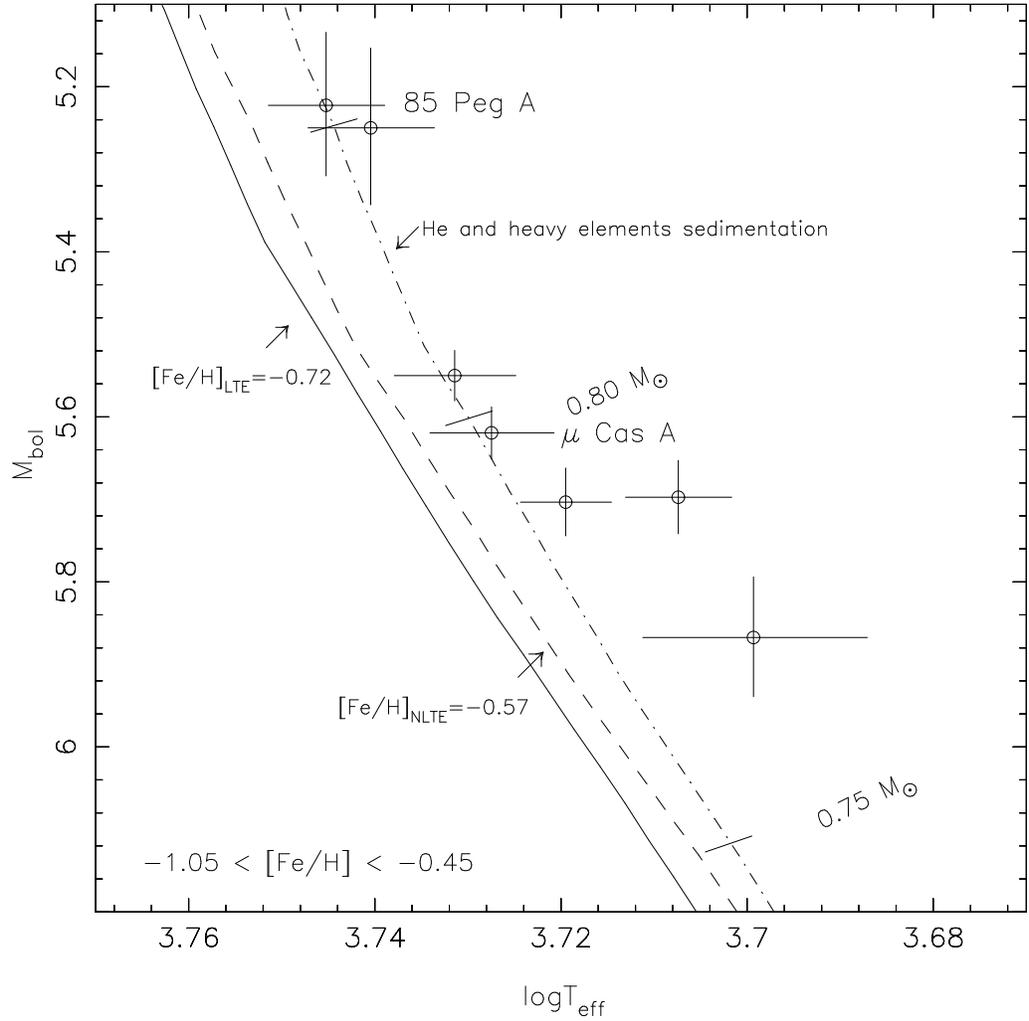}}

\caption{H-R diagram for the unevolved
moderately metal deficient stars of Figure~\ref{LebrF2} (mean
LTE metallicity $[\rm Fe/H]_{LTE}=-0.72$, mean non-LTE value
$[\rm Fe/H]_{NLTE}=-0.57$, see text). Full and dashed lines are standard
isochrones (10 Gyr) computed with, respectively, the $[\rm Fe/H]_{LTE}$
and $[\rm Fe/H]_{NLTE}$ values. The dot-dashed isochrone (10 Gyr) includes
He and heavy elements sedimentation: at the surface it has
$[\rm Fe/H]_{NLTE}=-0.57$ but the initial \FeH~ was $\approx$-0.5.
(from Lebreton et al 1999).}
\label{LebrF3}
\end{figure}

\noindent STATISTICAL STUDIES. \ \  Complete H-R diagrams of stars of the solar
neighborhood have been constructed by adopting different selection
criteria, and have been compared to synthetic H-R diagrams based on
theoretical evolutionary tracks.

\begin{itemize}
\item Schr\"oder (1998) proposed diagnostics of MS overshooting
based on star counts in the different regions of the ``Hipparcos''
H-R diagram of stars in the solar neighborhood (d$<$50-100 pc).
In the mass range 1.2-2 \Msun, convective cores are small, and it
is difficult to estimate the amount of overshooting with isochrone shapes.
Schr\"oder suggested using the number of stars in the Hertzsprung gap,
associated with the onset of H-shell burning, as an
indicator of the extent of overshooting around 1.6 \Msun; the
greater the overshooting on the MS, the larger the
He-burning core, and in turn the longer the passage through
the Hertzsprung gap. Actual star counts favor an
onset of overshooting around $\sim$1.7 \Msun (no
overshooting appears necessary below that mass), which is broadly
consistent with other empirical calibrations (MS width,
eclipsing binaries), but finer quantitative estimates would require
more accurate observational parameters, mainly in \Teff~ and \Z.

\item Jimenez et al (1998) compared the red envelope of ``Hipparcos''
subgiants (\sigpi$<$0.15, $\sigma_{\rm(B-V)}<$0.02 mag)
with isochrones to determine a minimum age of the
Galactic disk of 8~Gyr, which is broadly consistent with ages obtained
with other methods (white-dwarf cooling curves,
radioactive dating, isochrones, or fits of various age-sensitive
features in the H-R diagram). The fit is still qualitative: the
metallicities of subgiants are
unknown because of the inadequacy of model atmospheres in that region.
For this reason, Jimenez et al investigated the isochrones
in other regions, MS and clump (He core burning).
They calculated the
variations with mass of the clump position for a range of
metallicities in the disk, and showed that stars with masses from
0.8 to 1.3 \Msun (ages from 2 to 16 Gyr) all occupy a
well-defined vertical branch, the red-edge of the clump. The
color of this border line is sensitive to metallicity, which
makes it a good metallicity indicator in old metal-rich populations.

\item Ng \& Bertelli (1998) revised the ages of stars of the
solar neighborhood and derived corresponding age-metallicity and
age-mass relations. Fuhrmann (1998) combined the [Mg/H]-[Fe/H]
relation with age and kinematical information to distinguish
thin and thick disk stars. Several
features seem to emerge from these studies: (1) no evident age-metallicity
relation exists for the youngest ($<8$ Gyr) thin-disk stars; some of them
are rather metal-poor, and super metal-rich stars
appear to have been formed early in the history of the thin disk;
(2) there is an apparent lack of stars in the age-interval 10-12~Gyr which is
interpreted by Fuhrmann as a signature of the thin-disk formation; and
(3) beyond 12~Gyr there is a slight decrease of metallicity with increasing age for
stars of the thick disk; some of them are as old as halo stars.
To assess these suggestions and to assist progress in the understanding of
the Galactic evolution scenario (see Fuhrmann 1998 for 
details), enlarged stellar samples and further
improvements on age determinations are of course required.
\end{itemize}

\noindent THE SUBDWARF/SUBGIANT SEQUENCE.\ \
Hipparcos provided the very first high-quality parallaxes for a
number of halo stars. Age determinations of the
local halo could be undertaken, as well as comparisons with
globular cluster sequences.

\begin{figure}
\resizebox{\hsize}{!}{\includegraphics{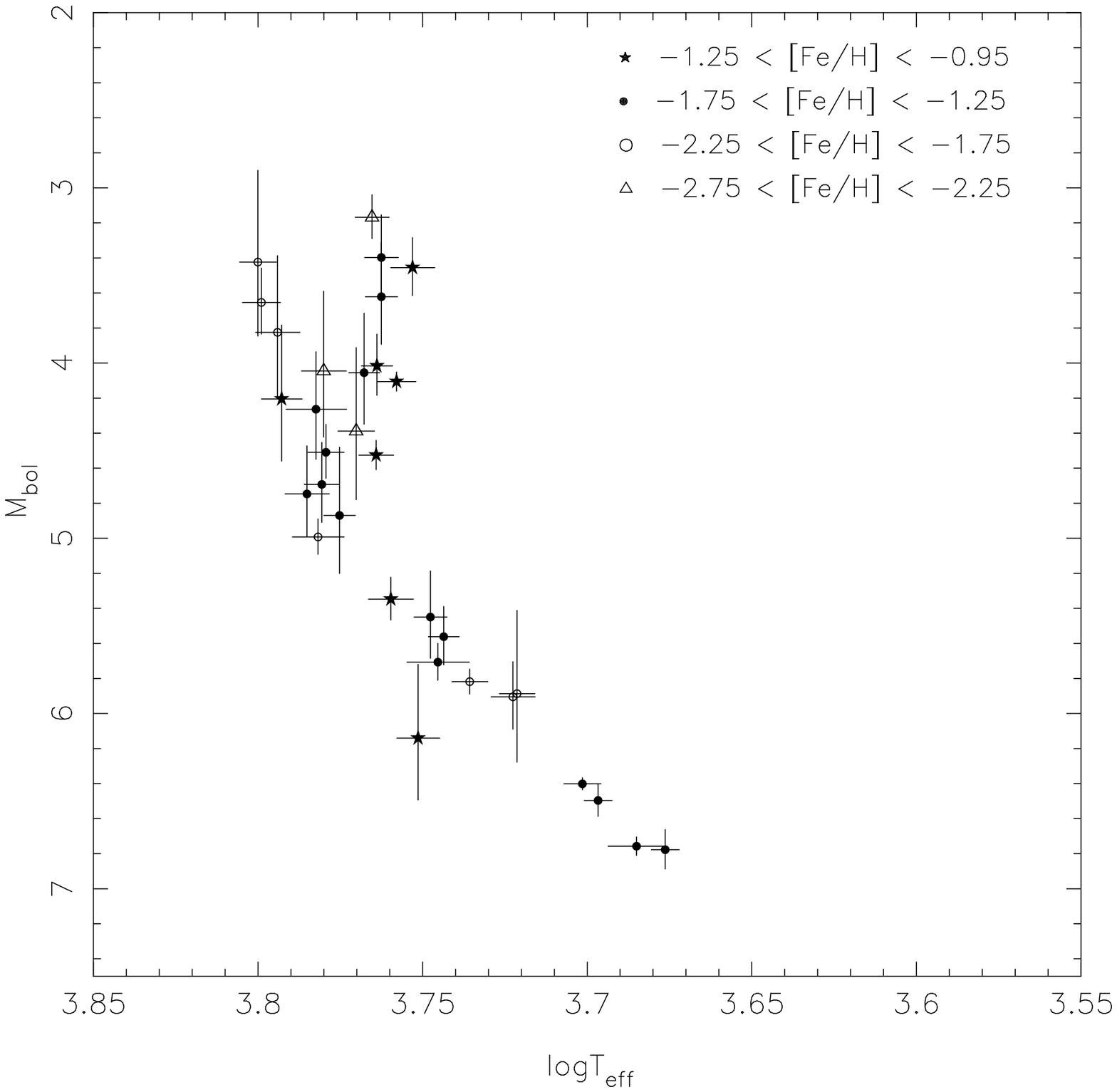}}

\caption{Hipparcos H-R diagram of the 32 halo stars with
\sigpi$<$0.22 (the parallax accuracies \sigpi are in the range 0.007-0.214).
Bolometric fluxes and effective temperatures are available from Alonso
et al's (1995, 1996a) works (see the caption for
Figure~\ref{LebrF1}). Resulting \sigMbol~ are in the range
0.03-0.48 mag. 
A bunch of subgiants emerges with an isochrone-like shape (from Cayrel et al
1997b).}
\label{LebrF4}
\end{figure}

Among a large sample of Population II Hipparcos halo subdwarfs,
Cayrel et al (1997b) extracted the best-known stars with
criteria similar to those adopted by Lebreton et al (1999) for
disk stars. Stars were corrected for reddening, excluding stars with
E(B-V)$>$0.05. Prior to Hipparcos, only 5 halo stars had
parallax errors smaller than 10\%; now there are
17, which represents sizeable progress. The halo stars
are plotted in Figure~\ref{LebrF4}; subdwarfs but also subgiants are present,
delineating an isochrone-like shape with a turn-off region.

\begin{figure}
\resizebox{\hsize}{!}{\includegraphics{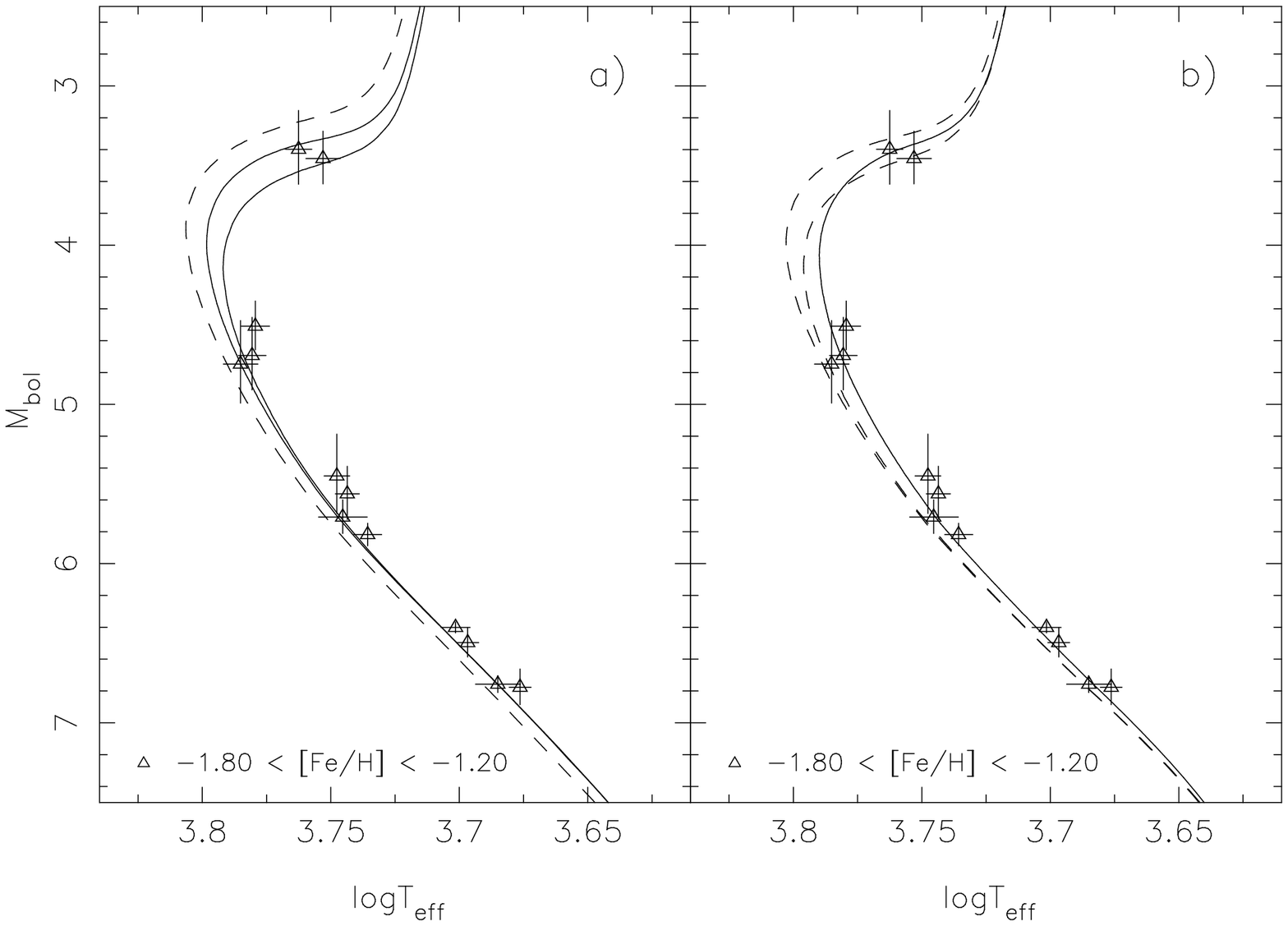}}

\caption{
Hipparcos H-R diagram of 13 halo stars with 
$[\rm Fe/H]_{LTE}=-1.5\pm0.3$ and \sigpi$<$0.12 
(the parallax accuracies \sigpi are in the range 0.01-0.12).
Bolometric fluxes and effective temperatures are available from Alonso
et al's (1995, 1996a) works (see the caption for
Figure~\ref{LebrF1}). Resulting \sigMbol~ are in the range
0.03-0.26 mag. All isochrones were kindly provided by DA Vandenberg.
Fig~\ref{LebrF5}a illustrates the effect of a non-LTE correction of +0.2 dex in \FeH~
as inferred from PA Bergbusch \& DA VandenBerg's (2000, in preparation)
models (see text): the dashed line is a standard isochrone
of 12 Gyr (\alFe=+0.3, \Y$\simeq$0.24) with \FeH=-1.54
(LTE value), and the full lines are isochrones with
\FeH=-1.31 (non-LTE value) of 12 Gyr (upper line)
and 14 Gyr (lower line). Fig~\ref{LebrF5}b illustrates
the effect of microscopic diffusion of He as inferred from
Proffitt's \& VandenBerg's (1991) models: isochrones 
(\FeH=-1.3 and [O/Fe]=0.55), of 12 Gyr with (full line)
and without (dashed-line; upper line 12 Gyr, lower 14 Gyr)
microscopic diffusion are plotted.
}

\label{LebrF5}
\end{figure}

To make a first estimate of the age of the local halo,
Cayrel et al kept 13 stars with the lowest error bars
and spanning a narrow metallicity range (\FeH=-1.5$\pm$0.3),
the most commonly found in the halo (Figure~\ref{LebrF5}).
They found that halo stars, like disk stars, are colder
than the theoretical isochrone corresponding to their metallicity.
The misfit was also noted by Nissen et al (1997) and Pont et
al (1997) in larger samples of halo stars.
The discrepancy amounts to 130 to 250~K depending on the
metallicity, and comparisons indicate that
it is independent of the particular set of isochrones used.
Again, non-LTE corrections leading to increased
\FeH-values ($\Delta$\FeH=+0.2 for \FeH$\sim$-1.5 according to
Th\'evenin \& Idiart 1999), added to the effects of microscopic diffusion,
can be invoked to reduce the misfit.
Figure~\ref{LebrF5}a compares Cayrel et al's sample with
standard isochrones by PA Bergbusch \&
DA VandenBerg (2000, in preparation), 
showing that the subdwarf main sequence cannot be reproduced
by isochrones computed with the LTE \FeH-value,
but increasing the metallicity (to mimic non-LTE corrections)
improves the fit. 
Figure~\ref{LebrF5}b compares the halo sequence with Proffitt
\& VandenBerg's (1991) isochrones that include He sedimentation.
Microscopic diffusion makes the isochrones redder,
modifies their shape, and predicts a lower turn-off luminosity: 
the best fit with the observed sequence is achieved
for an age smaller by 0.5-1.5 Gyr than that obtained without diffusion.
Models by Castellani et al (1997) show that, if sedimentation of metals is
also taken into account, including its effects on the matter opacity,
the isochrone shift is smaller than the shift obtained with He diffusion only.

Cayrel et al (1997b) and Pont et al (1997) estimated 
the local halo to be 12-16 Gyr old (from standard isochrones).
To improve the precision more stars with accurate parallaxes are required.
Subgiants are about 100 times rarer than subdwarfs, and we
have only two subgiants with \sigpi$<$12.5\% (and no subgiant with \sigpi$<$5\%).
After Hipparcos the position of the subgiant branch is
still poorly determined, which limits the accuracy on the
age determination of the halo stars.

\begin{figure}
\resizebox{\hsize}{!}{\includegraphics{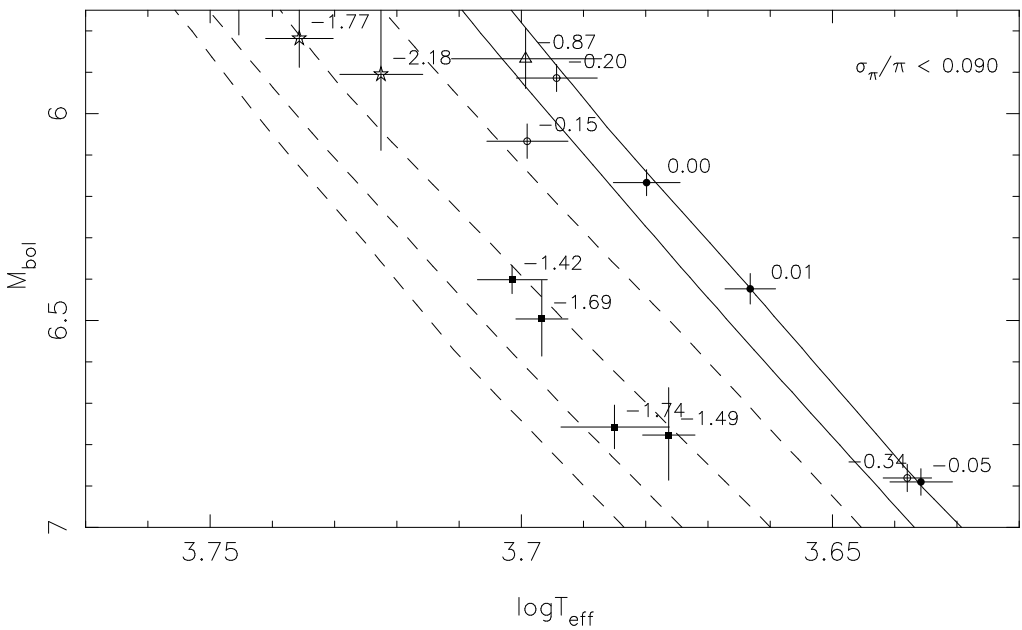}}

\caption{``Hipparcos'' H-R diagram of non-evolved stars
with \sigpi$<$0.10. Each star is labeled with its \FeH-value.
Standard isochrones are plotted with, from left to right,
\FeH=-2.0, -1.5, -1.0, -0.5, 0.0, 0.3 (from Lebreton 2000).}
\label{LebrF6}
\end{figure}

\noindent THE ZAMS POSITIONS. \ \
The sample made of Hipparcos disk and halo stars spans
 the whole Galactic metallicity range.
Figure~\ref{LebrF6} shows the non-evolved stars (\Mbol$>$5.5)
of Figure~\ref{LebrF1} and Figure~\ref{LebrF4} along with
standard isochrones of various metallicities and solar-scaled
helium (\dydzsun=2.2). It allows a discussion of the
position of the zero age main sequence (ZAMS) as a function of
metallicity and implications for the unknown helium abundances.

\begin{itemize}
\item {\sl MS width.} Although stars generally do not lie where
predicted, in particular at low metallicities, the
observational and theoretical MS widths are in
reasonable agreement for \dydz=2.2.
This qualitative agreement is broadly consistent with 
\dydz ratio of $\simeq$3$\pm$2 derived from similar measures of
the lower MS width by Pagel \& Portinari (1998) and the lower limit
\dydz$\gtrsim$2 obtained by Fernandes et al (1996) from pre-Hipparcos
MS. It also agrees with extragalactic determinations (see Izotov et al
1997) or nucleosynthetic predictions.

\item {\sl Helium abundance at solar metallicities.} It can be noted from
Figure~\ref{LebrF6} that there are 4 stars with Fe/H close to
solar on the \FeH=0.3 isochrone. Non-LTE \FeH-corrections are
negligible at solar metallicity. Microscopic diffusion may produce
a shift in the H-R diagram: for a 0.8 \Msun~ star of solar Fe/H at
5~Gyr the shift is small and comparable to the observational
error bars (but it increases with age). These disk stars are not
expected to be very old and the shift could instead indicate that
their He-content is lower than the solar-scaled value.
Calibration of individual objects and groups
with metallicities close to solar indicate an increase of helium
with metallicity corresponding to \dydz$\simeq$2.2 from the Sun (Lebreton et
al 1999) and \dydz$\simeq$2.3$\pm$1.5 from visual binaries 
(Fernandes et al 1998) but exceptions are found, such as in the (rather young)
Hyades which, although metal-rich (\FeH=0.14), appear to have a
solar or even slightly sub-solar helium content
with \dydz$\simeq$1.4 (Perryman et al 1998). Going into finer
resolution would clearly require more complete data including masses
for enlarged samples of non-evolved stars.

\item {\sl Position of metal-deficient stars.}
Very few metal-deficient stars
have accurate positions in the non-evolved part of the H-R diagram:
a gap appears for \FeH$\in$[-1.4,-0.3] and only 4 subdwarfs are found
below \FeH$\sim$-1.4. The empirical dependence of the ZAMS location with
metallicity is impossible to establish for these stars, which are
expected to have practically
primordial helium contents. This adds to difficulties in
estimating the distances of globular clusters (Eggen \& Sandage 1962; Sandage 1970, 1983;
Chaboyer et al 1998).
\end{itemize}

\subsection{Stars in Open Clusters}
\label{OC}

Hipparcos observed stars in all open clusters closer than 300 pc
and in the richest clusters located between 300 and 500 pc providing
valuable material for distance scaling of the Universe and
for studies of kinematical and chemical evolution of the
Galaxy.
The absolute cluster sequences in the H-R diagram
may be constructed directly from Hipparcos distances independently
of any chemical composition consideration.
Each sequence covers a large range of stellar masses and contains
stars which can reasonably be considered to be born at the same
time with similar chemical composition. Several clusters provide
tests of the internal structure models for a wide range of
initial parameters, in particular for different metallicities.

\noindent THE HYADES. \ \ Obtaining high-quality astrometric data for
the Hyades has been crucial, for it is the nearest rich star
cluster, used to define absolute magnitude calibrations and
the zero-point of the Galactic and extragalactic distance scales.
Individual distances (mean accuracy of 5\%)
and proper motions were given by Hipparcos, providing a
consistent picture of the Hyades distance,
structure and dynamics (Perryman et al 1998).
The recent determinations of the Hyades distance modulus $(m-M)$
are all in very good agreement while the internal accuracy was
largely improved with Hipparcos:

\begin{itemize}
\item ground-based: $m-M= 3.32\pm0.06$ mag (104 stars, van Altena et al 1997b)
\item \HST: $m-M =3.42\pm0.09$ mag (7 stars, van Altena et al 1997a)
\item Hipparcos: $m-M= 3.33\pm0.01$ mag (134 stars within 10\ pc of
the cluster center, Perryman et al 1998)
\item statistical parallaxes based on Hipparcos proper motions:
$m-M=3.34\pm0.02$ mag (43 stars, Narayanan \& Gould 1999a who also
showed that the systematic error on the parallaxes toward the Hyades is
lower than 0.47 mas).
\end{itemize}

\begin{figure}
\resizebox{\hsize}{\hsize}{\includegraphics{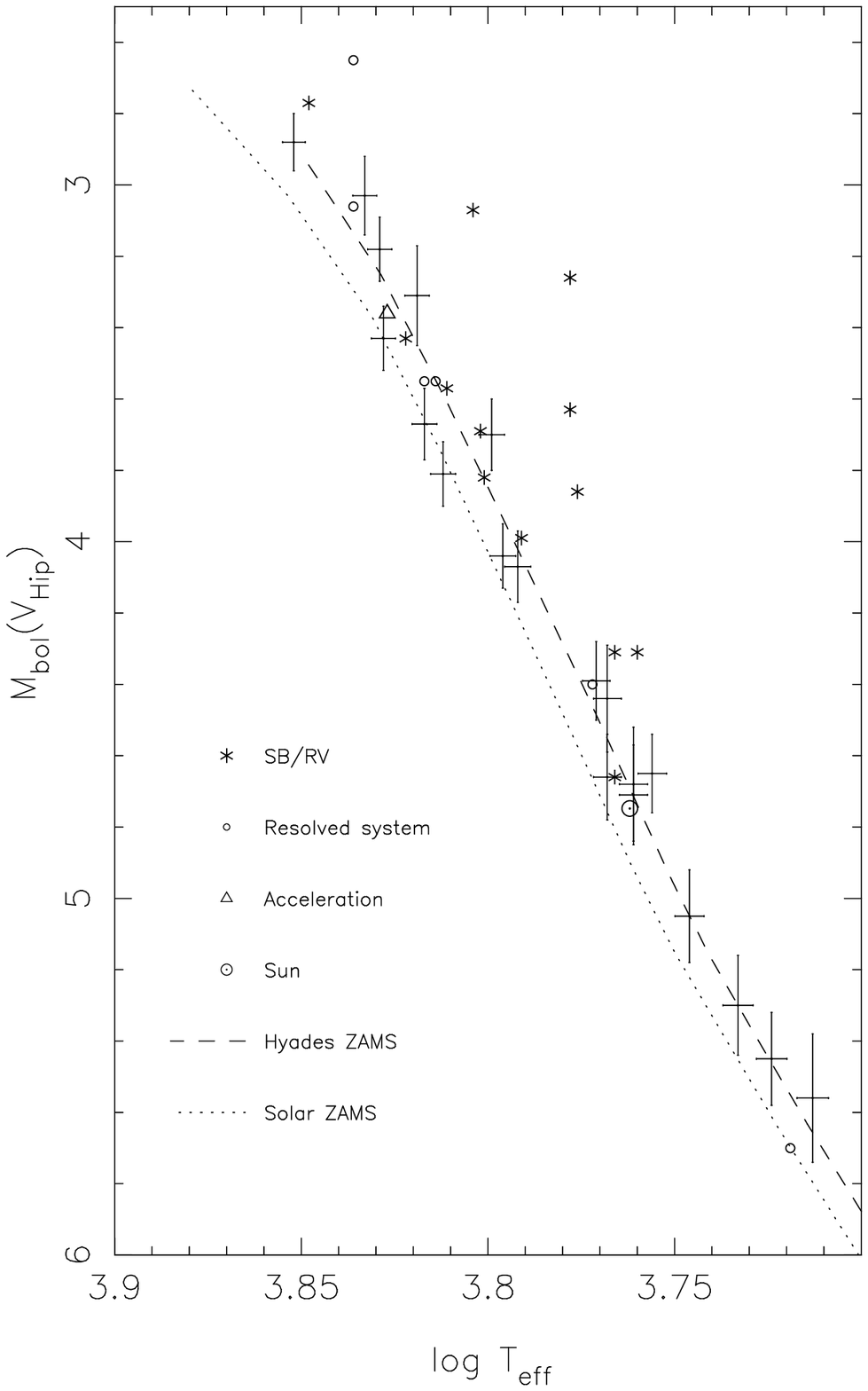}}

\caption{Hipparcos H-R diagram for 40 selected low MS stars in the
Hyades. The mean \FeH~ is 0.14$\pm$ 0.05. Stars with error bars are not suspected to be
double or variable. Internal errors on \Teff~ are in the range 50-75\ K.
Double or variable stars are also indicated: objects resolved by
Hipparcos or known to be double systems are shown as circles,
triangles denote objects with either detected photocentric
acceleration or objects possibly resolved in photometry, and `$\ast$' means 
spectroscopic binary or radial velocity variable.
Theoretical ZAMS loci are given for the
Hyades (dashed line, \Y=0.26 \Z=0.024) and
solar (dotted line, \Y=0.266 \Z=0.0175) chemical compositions
(from Perryman et al 1998).}
\label{LebrF7}
\end{figure}

\begin{figure}
\resizebox{\hsize}{!}{\includegraphics{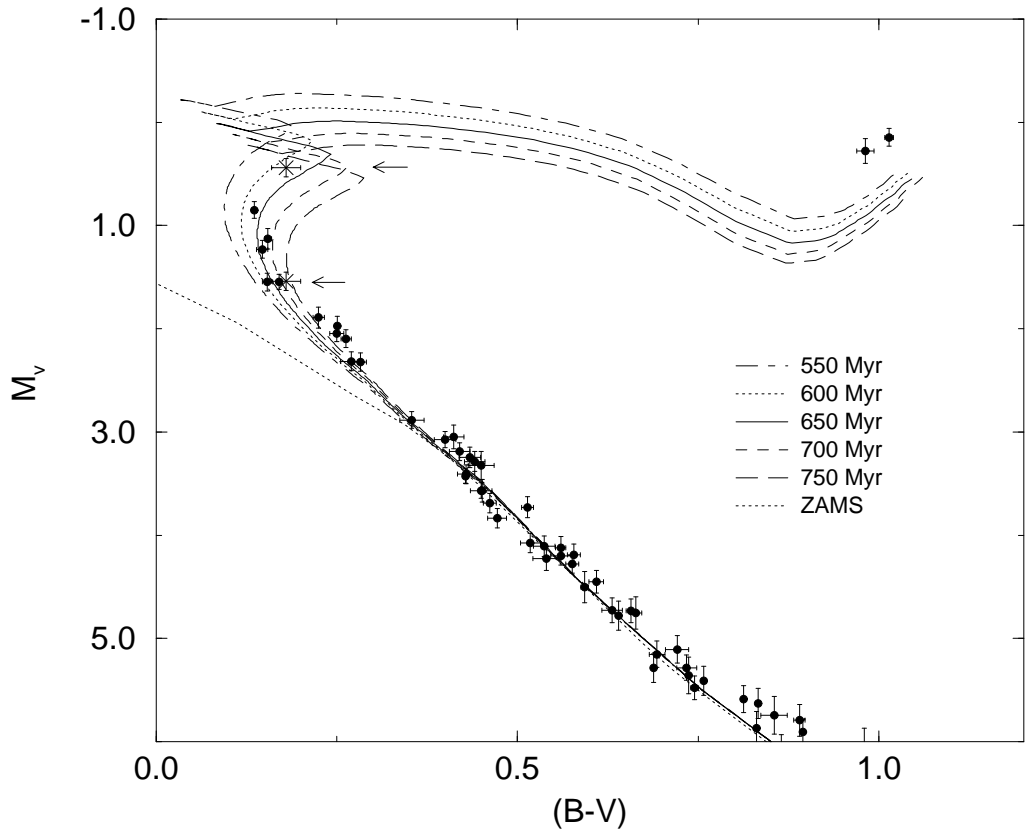}}

\caption{Hipparcos C-M diagram of the Hyades. V and B-V values are from the
Hipparcos catalogue ($\sigma_{\rm(B-V)}<$ 0.05 mag).
The loci of ZAMS and theoretical isochrones calculated
with overshooting (\aov=0.2) are indicated. Arrows indicate the position of
the components of the binary system $\theta^2$ Tau used
for the age determination (from Perryman et al 1998).}
\label{LebrF8}
\end{figure}

\begin{figure}
\resizebox{\hsize}{!}{\includegraphics{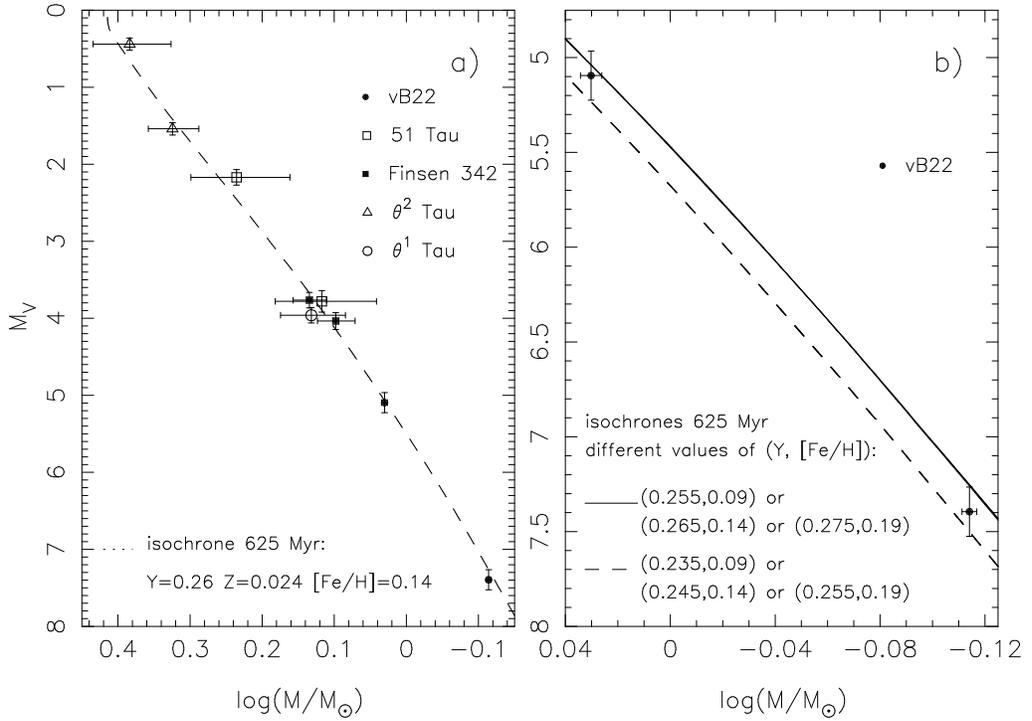}}

\caption{Figure~\ref{LebrF9}a: the Hyades empirical and
theoretical mass-luminosity relations.
Masses are from Peterson \& Solensky (1988, vB22),
Torres et al (1997a, b, c, Finsen 342, $\theta^{1}$, $\theta^{2}$ Tau)
and Söderhjelm (1999, 51 Tauri from Hipparcos data).
The isochrone is for 625 Myr, taken from the same calculations used 
in Figure~\ref{LebrF8}.
Figure~\ref{LebrF9}b illustrates how the precise positions of the
members of vB22 allow to discriminate between different (\Y, \FeH)
values.} 
\label{LebrF9}
\end{figure}

Greatly improved precision is seen in the H-R diagrams built with
Hipparcos data combined with the best ground-based observations
(Perryman et al 1998):
\begin{itemize}
\item Figure~\ref{LebrF7} shows 40 stars with \Teff~ and
\FeH=0.14$\pm$0.05 from detailed spectroscopic analysis
delineating the lower part of the observational MS of the cluster (Cayrel de
Strobel et al 1997a).

\item Figure~\ref{LebrF8} is the whole H-R diagram
in the (\Mv, B-V) plane for 69 cluster members. Known or
suspected binaries, variable stars, and rapid rotators have been
excluded (Perryman et al 1998). Also, Dravins et al (1997) 
derived dynamical parallaxes for the Hyades
members from the relation between the cluster space motion, 
the positions and the projected
proper motions; these parallaxes are more precise (by a factor of
about 2) than those directly measured by Hipparcos, yielding in
turn a remarkably well-defined MS sequence in the H-R diagram,
narrower than that given in Figure~\ref{LebrF8} (see Figure 2 in
Dravins et al 1997). 

\item Figure~\ref{LebrF9} shows the empirical mass-luminosity
(M-L) relation drawn from the (very accurate) masses of 5 binary
systems (see caption). Nine of the stars are MS stars.
\end{itemize}

Comparisons with theoretical models yield some of the
cluster characteristics (Lebreton et al 1997a, Perryman
et al 1998, Lebreton 2000):

\begin{enumerate}
\item The comparison of the lowest part of the MS (Figure~\ref{LebrF7}),
representing the non-evolved stars, with theoretical ZAMS corresponding to the
mean observed \FeH~ yields the initial cluster helium
content $Y_H$=0.26$\pm$0.02 and metallicity $Z_H$=0.024$\pm$0.04.
Metallicity is the dominant source of the uncertainty on \Y .

\item The comparison of the whole observed sequence with model isochrones
yields the cluster age. Figure~\ref{LebrF8} shows that the optimum
fit is achieved with an isochrone of $625\pm50$ Myr,
$Y_H$=0.26, $Z_H$=0.024 and including overshooting. The turn-off
region (which in the Hyades corresponds to the instability strip of 
$\delta$ Scuti stars)
is rather well represented by the 625 Myr isochrone (see also
Antonello \& Pasinetti Fracassini 1998).
The quoted uncertainty on age only includes the contribution from
visual fitting of the isochrones. Additional errors on age result from
unrecognized binaries, rotating stars, color calibrations and
bolometric corrections, and from theoretical models
in particular through the parameterization of overshooting (Lebreton et al
1995). It is therefore reasonable to give an overall
age uncertainty of at least 100 Myr.

\item In Figure~\ref{LebrF9}a the observed M-L relation is compared
with the theoretical isochrone of 625 Myr, $Y_H=0.26$, $Z_H= 0.024$,
showing an excellent agreement. The lower part of the relation is
defined by the very accurate masses of
the two components of vB22. 
This system gives additional
constraints on the $Y_H$-value derived from ZAMS calibration.
Figure~\ref{LebrF9}b illustrates how the positions of the two
vB22 components may be used to constrain $Y_H$ in the whole
metallicity range allowed by observations, \FeH=0.14$\pm$0.05 
(see also Lebreton 2000).
\end{enumerate}

Furthermore, in the turn-off region of
the Hyades, 5 $\delta$ Scuti stars are found that are quite rapid
rotators ($v_e \sin i$ in the range 80-200 $\rm km.s^{-1}$,
see Antonello \& Pasinetti Fracassini 1998). From
the measurement and analysis of their oscillation frequencies and
the identification of the corresponding modes by
means of models (of same age and chemical composition),
we should be able to derive the inner rotation profile and
learn about the size of convective cores and transport
processes at work in the interiors (Goupil et al 1996, Michel et
al 1999). For instance, the rotation profile is related to the
redistribution of angular momentum by internal motions which
could be generated by meridional circulation and shear turbulence in a
rotating medium (see Zahn 1992). On the other hand, such motions might induce
internal mixing, and as shown by Talon et al
(1997), in the H-R diagram rotational effects ``mimic'' overshooting
(for instance, in a star of 9 \Msun, a rotational velocity of
$\rm \sim100\ km.s^{-1}$ is equivalent
to an overshooting of \aov$\sim$0.2).

The study and intercomparison of accurate observations of the 
non-pulsating and pulsating stars located in the instability strip
should clearly provide deeper insight into the internal
structure and properties of stars of the Hyades cluster.
However, such analysis has to integrate the various complications
related to rotation, such as the displacements in any
photometric H-R diagram by amounts depending on the equatorial
velocity and inclination (Maeder \& Peytremann 1972,
P\'erez-Hern\'andez et al 1999)
or the splitting of oscillation frequencies,
which has to be considered in the mode identification.

\noindent THE PLEIADES AND OTHER OPEN CLUSTERS. \ \
The membership of stars in nine clusters closer than 300 pc was
carefully assessed by van Leeuwen (1999a) and Robichon et al
(1999a). Robichon et al also studied nine rich clusters  within 500
pc with more than 8 members and 32 more distant clusters.
For clusters closer than 500 pc, the accuracy on the mean parallax is in
the range 0.2-0.5 mas and the accuracy on the mean proper motions is of
the order of 0.1 to 0.5 mas per year. Results from the two
groups are in very good agreement. Platais et
al (1998) looked for (new) star clusters in Hipparcos data
and found one new, a nearby cluster in Carina (d=132 pc) with 7
identified members.

In order to obtain an optimal mean parallax with correct error estimates,
van Leeuwen (1999a) and Robichon et al (1999a) worked with the
Hipparcos intermediate data corresponding to each cluster, 
parallax and proper motion of the cluster center, and
position of each cluster member, instead of making a straight 
average of the parallaxes of the
cluster members. Stars in open clusters are located
within a few degrees on the sky and hence
were often observed in the same field of view of the satellite.
A combined solution can be obtained from intermediate data,
which allows angular correlations to be taken
into account and the resulting parallax errors to be minimized 
(van Leeuwen \& Evans 1998; van Leeuwen 1999a, b; Robichon et al 1999a).

Mermilliod et al (1997), Robichon et al (1997), and Mermilliod (2000)
compared the sequences of the various clusters in C-M
diagrams derived from different photometric systems, and found
puzzling results that are at odds with the common idea that
differences in metallicity fully explain the 
relative positions of the non-evolved parts of the MS of
different clusters:

\begin{itemize}
\item Some clusters have different
metallicities but define the same main sequence in the (\Mv, B-V) plane
(Praesepe, Coma Ber, \alf Per, Blanco 1). For instance,
Coma Ber has a quasi-solar metallicity while its sequence is similar
to that of the Hyades, or of the metal-rich Praesepe.

\item Some clusters sequences (Pleiades, IC 2391 and 2602)
are abnormally faint with respect to others, for instance Coma Ber.
The metallicity of the Pleiades as determined from
spectroscopy is almost solar, and similar to that of Coma Ber,
but the Pleiades sequence lies (unexpectedly) $\sim$~0.3-0.4 mag below the Praesepe,
Coma Ber, or Hyades sequence.

\item Van Leeuwen (1999a, b) even suggested a possible (although
unexpected) correlation between the age of a cluster 
and its position in the H-R diagram. 
\end{itemize}

Prior to Hipparcos, precise trigonometric parallaxes had not been
obtained for clusters except the Hyades. Distances to open clusters
were evaluated through the main sequence fitting technique:
the non-evolved part of the (observed) cluster sequence was
compared to the non-evolved part of the (absolute) lower MS
(ZAMS) of either (1) theoretical isochrones, (2) field stars or
(3) Hyades after a possible correction of chemical composition differences.
The magnitude differences between absolute and apparent ZAMS
directly yielded the distance modulus of the cluster. 

The Hipparcos distances to the 5 closest open clusters
(Hyades, Pleiades, \alf Per, Praesepe, and Coma Ber)
can be compared to those recently derived from MS fitting by
Pinsonneault et al (1998); they compared theoretical isochrones,
translated into the C-M plane by means of Yale color calibrations,
to observational data both in the (\Mv, B-V) and (\Mv, V-I) planes.
The B-V color indice is more sensitive to metallicity than V-I
(Alonso et al 1996b), so Pinsonneault et al derived as
a by-product the value of the metallicity that gives
the same distance modulus in the two planes and compared it to
spectroscopic determinations. They judged their distance modulii
to be in good agreement with Hipparcos results except for the
Pleiades and Coma Ber. For Coma Ber,
the problem could result from the old VRI colors used.
For the Pleiades the discrepancy with
Hipparcos amounts to 0.24 mag, and the \FeH-value derived
from MS fitting in the two color planes agrees
with the spectroscopic determination of Boesgaard \&
Friel (1990), \FeH= -0.034$\pm$0.024, although values in the range
-0.03 to +0.13 can be found in the literature. In fact, with
that metallicity the Hipparcos sequence of the
Pleiades could be reproduced by classical theoretical models,
provided they have a high helium content. The
exact value depends on the model set and its input physics:
Pinsonneault et al found \Y=0.37, Belikov et al (1998) found \Y=0.34 but
for \FeH=0.10 and I find Y $\sim$ 0.31. In any case, high helium
content is only marginally
supported by observations (Nissen 1976). Pinsonneault et al examined
other possible origins of the discrepancy
(erroneous metallicity, age-related effects, reddening)
and concluded that none of them is likely to
be responsible for the Pleiades discrepancy. 

In parallel, Soderblom et al (1998) looked for
young solar-type stars appearing as (anomalously) faint as
the Pleiades. They found 50 field stars expected to be young
(i.e. showing activity from Ca II H and K lines),
but none of them lies significantly below the ZAMS. They also examined 
the subluminous stars observed by Hipparcos: they chose six stars
among those lying well below the ZAMS, measured their
spectroscopic metallicities, and found them to be
metal-deficient with respect to the Sun with, in addition, kinematics
typical of stars of a thick disk or halo population.

Soderblom et al and Pinsonneault et al concluded that, taking the
Hipparcos results for the Pleiades at face value, it would be
abnormal not to find stars similar to the Pleiades in the
field. They inferred that the distance obtained from multi-color
MS fitting is correct and accurate to about
0.05 mag, and concluded that the distance to Pleiades
obtained from the analysis of Hipparcos data is
possibly wrong at the 1 mas
level, which is greater than the mean random error.
They invoked statistical correlations between right ascension and parallax
($\rho_{\alpha \cos \delta}^\pi$) arising from the non-uniform distribution of
Hipparcos observations over time (and in turn along the
parallactic ellipse) which affects all stars, including clusters.
Pinsonneault et al noted that in the Pleiades the brightest
stars (1) are highly concentrated near the cluster center and are therefore
subject to spatial correlations which gives them nearly the same
parallax, (2) have smaller $\sigma_\pi$ than fainter stars which
gives them more weight in the mean parallax, and (3) are those which
have the highest values of $\rho_{\alpha \cos \delta}^\pi$ and
also the highest parallaxes in the Hipparcos Catalogue.
They suggest that the ``true'' parallax
(close to that obtained through MS fitting)
is obtained if the brightest stars with high
$\rho_{\alpha \cos \delta}^\pi$ are excluded from the calculation.

Narayanan \& Gould (1999b) determined the parallaxes of the 
Pleiades stars by means of Hipparcos proper motions. The
resulting distance modulus has a rather large error bar
($m-M=5.58 \pm 0.18$ mag), but it is in disagreement 
with that derived directly from Hipparcos parallaxes 
($m-M=5.36 \pm 0.07$ mag), and in agreement
with that obtained through MS fitting ($m-M=5.60 \pm 0.05$ mag).
Narayanan \& Gould also argue that the
differences between the Hipparcos trigonometric parallaxes and the
parallaxes derived from Hipparcos proper
motions reflect spatial correlations over small angular scales 
with an amplitude of up to 2 mas.

Robichon et al (1999a, b) and van Leeuwen (1999b) have
subsequently derived more reliable distance estimates to these 
clusters and performed tests that do not support Pinsonneault et al's
conclusion. The difference between
Hipparcos and MS fitting distance moduli is small for the
Hyades (0.01 mag), whereas for other clusters it ranges from -0.17 mag (\alf
Per) to +0.24 mag (Pleiades). In fact, except for the Hyades, the difference is
always larger than the error on MS fitting distance modulus (0.05
mag). Robichon et al showed that while the solution proposed by Pinsonneault et al
improves the situation for the Pleiades, it would introduce
new difficulties for Praesepe. By means of Monte-Carlo
simulations of the Pleiades stars, they showed
that the mean value of the Pleiades parallax
does not depend on the correlations $\rho_{\alpha \cos \delta}^\pi$. 
They also carefully examined distant
stars and clusters with high $\rho_{\alpha \cos \delta}^\pi$.
Through these tests, Robichon et al made the Hipparcos distance to Coma Ber or
Pleiades more secure, and did not find any obvious bias
on the parallax resulting from a correlation between
right ascension and parallax, either for stars within a
small angular region or for the whole sky. 

On the other hand, distances from MS fitting could be
subject to higher error bars than quoted by Pinsonneault et al.
They depend on reddening and on transformations from the
(\Mbol, \Teff) to the C-M plane if
theoretical ZAMS are used as reference (or on metallicity corrections
if empirical ZAMS are compared). Robichon et al (1999b)
compared solar ZAMS from Pinsonneault et al to those I
calculated both in the theoretical and in the (\Mv, B-V) planes.
They showed that while the two ZAMS are within 0.05 mag in the
theoretical H-R diagram, they differ by 0.15-0.20 mag in the
range B-V=0.7-0.8 in the (\Mv, B-V) plane, simply because
different C-M transformations have been applied. 
Also, MS fitting often relies on rather old and inhomogeneous
color sources (in the separate Johnson and Kron-Eggen RI systems) requiring
transformations to put all data on the same (Cape-Cousins system) scale.
It would therefore be worthwhile to verify the quality and precision
of these data by making new photometric measurements of cluster stars.

Let us come back to  the difficult question of metallicity.
As pointed out by Mermilliod et al (1997), photometric and
spectroscopic approaches may produce quite different results.
Metallicities have been derived recently by M Grenon (1998, private
communication) from large sets of homogeneous observations
in the Geneva photometric system. He obtained \FeH=-0.112$\pm$0.025 for
the Pleiades (quite different from published spectroscopic values)
and \FeH=0.170$\pm$0.010
and 0.143$\pm$0.008 for Praesepe and the Hyades respectively (both in
agreement with spectroscopy). The observed cluster sequences obtained
with Hipparcos distances for the three clusters can be
roughly reproduced by theoretical models computed with the
photometric metallicities (and allowing for small variations of
the helium content around the solar-scaled value) and transformed
to the C-M plane according to the Alonso et al (1996b) and Bessell et al
(1998) calibrations (Robichon et al 1999b).

In conclusion, we point out that a detailed study of the fine
structure of the H-R diagram of the Pleiades (and other clusters) requires
supplementary observations (colors and abundances) and further
progress in model atmospheres. Today there is no obvious solid
argument against the published Hipparcos distances.
In order to identify and understand the remaining discrepancies
with stellar models, the entire set of observed clusters has to be
considered (van Leeuwen 1999a). Furthermore, not only the positions of the
sequences in the H-R diagram but also the density of stars
along them have to be intercompared.
For instance, the luminosity function of young clusters exhibits
a particular feature (local peak followed by dip) that is
interpreted as a signature of pre-MS stars and might provide information on the
initial mass function and stellar formation history
(see the study of the Pleiades by Belikov et al 1998).
On the other hand, since the error bars on
luminosity are now small with respect to errors
on color indices, stronger constraints are expected from the
mass-luminosity relation, as in the Hyades.
Observations of binaries in clusters are urgently needed
and there is hope to detect them in the future, for, as pointed out by Soderblom et
al, the difficult detection and measurement of visual binary orbits
in the Pleiades is within the capabilities of experiments on
board \HST.

\section{RARE, FAINT, SPECIAL, OR INACCESSIBLE OBJECTS}
\label{rare}

\subsection{Globular Clusters Through Halo Stars}
\label{GC}

Globular clusters were beyond the possibilities of Hipparcos, but
the knowledge of distances to nearby subdwarfs gave distance estimates
to a few of them through the MS
fitting technique (Sandage 1970; Reid 1997, 1998; Gratton et
al 1997; Pont et al 1997, 1998; Chaboyer et al 1998), comparing
the non-evolved part of the (absolute) subdwarf main-sequence
to the non-evolved part of the (observed) globular cluster sequence.
Although simple, the technique has to be applied with caution:

\begin{itemize}
\item Only halo subdwarfs and globular clusters
with the most precise data should be retained.
Abundances should be accurate and on a consistent scale.
Globular cluster abundances are usually determined only for
giants, while recent preliminary values have been obtained for
subgiants in M92 (King et al 1998). Abundance comparisons
between (1) field and cluster stars and (2) dwarfs and giants
have shown sometimes puzzling differences (King et al
1998, Reid 1999). Questions have been raised as to whether they are
primordial or appear during evolution, but definite answers clearly
require better spectra for all types of stars as well as spherical
model atmospheres with better treatment of convection.
The globular cluster sequence should be determined
from good photometry well below the MS turn-off, and the 
correction for interstellar reddening has to be well estimated.
Very few halo stars have parallaxes accurate enough to
fix precisely the position of the ZAMS (Section~\ref{nearby}).

\item Biases (see e.g. Lutz \& Kelker 1973; Hanson 1979; Smith 1987) 
resulting from the selection of the sample in apparent
magnitude, parallax, and metallicity, have to be corrected
for (Pont et al 1998, Gratton et al 1997); alternatively, samples
free of biases must be selected, which implies retaining the
very nearby stars with highly accurate parallaxes (Chaboyer et al
1998, Brown et al 1997).

\item Globular cluster sequence and halo sequence should (ideally) 
have similar {\sl initial} chemical compositions. Because of
the small number of subdwarfs in each
interval of metallicity, it is not possible to properly establish
the variation of the observed ZAMS position with metallicity, and
to correct for chemical composition differences between globular
clusters and subdwarfs empirically. Chaboyer et al
(1998) found it safer to limit the method
to globular clusters that have their equivalent in the field
with the same \FeH~ and \alFe~ content.
Gratton et al (1997) and Pont et al (1998) applied theoretical
color corrections to the subdwarf data
to account for metallicity differences with globulars.
In addition, element sedimentation might
introduce further difficulties, as already mentioned 
in Section~\ref{nearby}. As pointed out by Salaris \& Weiss
(2000), the present surface chemical composition of field subdwarfs
no more reflects the initial one if microscopic diffusion has
been efficient during evolution, while, in globular cluster
giants which have undergone the first dredge-up, the chemical
abundances have been almost restored to the initial ones.

\item Unresolved known or suspected binaries can introduce
errors in the definition of the ZAMS position. Chaboyer et al and
Gratton et al excluded them whereas Pont et al applied an
average correction of 0.375 mag on their position, a procedure that has been
criticized by Chaboyer et al and Reid (1998).

\item Evolved stars have to be excluded (there is no certainty that globular
clusters and halo dwarfs have exactly the same age).
From theoretical models it is estimated that stars fainter
than \Mv$\simeq$ 5.5 are essentially unevolved.

\end{itemize}

The number of globular clusters studied by the different authors
varies because of the different
criteria and techniques chosen to select the subdwarfs samples.
Nevertheless they all agree on the general conclusion that globular
cluster distances derived from MS fitting are larger
by $\sim$5-7\% than was previously found. Chaboyer (1998)
calculated an average of distances to globular clusters obtained with
different methods (MS fitting, astrometry, white dwarf sequence fitting,
calibration of the mean magnitudes of RR Lyrae stars in the Large
Magellanic Cloud, comparison with theoretical models of
horizontal branch stars, statistical parallax absolute magnitude
determinations of field RR Lyrae from the Hipparcos proper
motions, etc) and noted that the distance scale is larger (by
0.1 mag) than his pre-Hipparcos reference. 

It is worth pointing out that the statistical parallax method alone
favors a shorter distance scale
(by $\approx$0.3 mag with respect to MS fitting result). As
reviewed by Layden (1998) and by Reid (1999), the statistical
parallax method was applied by independent groups who found
concordant results. However, the absolute magnitudes $\rm M_V(RR)$
of the halo RR Lyrae derived from the 
statistical parallax method (on the basis of Hipparcos proper
motions and of radial velocities) are also $\approx$ 0.3 mag fainter than
the magnitudes obtained through a method directly based on
Hipparcos parallaxes (Groenewegen \& Salaris 1999); these latter
being in turn in good agreement with the MS fitting result. As
discussed thoroughly by Reid (1999), there are several difficulties 
related to the $\rm M_V(RR)$-calibration and to its comparison with other
distance calibrations that still hinder the
coherent and homogeneous understanding of the local distance scale.

In Caputo's (1998) and Reid's (1999) reviews the ages of
globulars are discussed.
After Hipparcos, ages of globular clusters are reduced
by typically 2-3 Gyr, because of both larger distances and
improvements in the physics of the models, mainly in the equation of
state and in the consideration of microscopic diffusion. Present ages
are now in the range 10-13~Gyr, which can be compared
to the previous interval of 13-18 Gyr (see 
VandenBerg et al's review, 1996).

Chaboyer et al (1998) claimed that the (theoretical) absolute magnitude
\Mv~ and lifetimes of stars at the MS turn-off (T-O) in
globular clusters are now well understood, since the physics involved
is very similar to that of the Sun, which is in turn well constrained by
seismology. In particular, \Mv(T-O) is quite insensitive to
uncertainties related to model atmospheres or convection
modeling (see also Freytag \& Salaris 1999). Chaboyer et al
suggested that no significant changes
(more than $\sim$5\%) in the derived ages of globular clusters
are expected from future improvements in stellar models.
Conversely, distance and abundance determinations are far from
definite, and the quasi-verticalness of isochrones in the T-O
region makes the determination of \Mv(T-O) difficult (see
Vandenberg et al 1996). Further revision of the ages is therefore
not excluded. Also a global agreement between an entire globular 
cluster sequence and the corresponding model isochrone is far from being reached.

{\sl Age of the Universe.} The ages of the oldest objects
in the Galaxy, the most metal-poor halo or globular cluster stars,
provide a minimum value for the age of the Universe $\rm T_U$.
Globular cluster ages (10-13~Gyr) from comparisons of isochrones with
observed \Mv(T-O) are presently the most reliable,
but two independent methods look promising:
\begin{itemize}
\item Thorium (half-life 14.05~Gyr) has been detected and
measured by Sneden et al (1996) in an ultra-metal-poor giant, too
faint for observation by Hipparcos, and
the star radioactive decay age is estimated to be 15$\pm$4~Gyr.
In the future, such observations of more stars and the possible
detection of Rhenium and Uranium could provide strong constraints
for $\rm T_U$.
\item  Observations of (faint) white dwarfs (WD) in globular
clusters are now within reach of experiments on board \HST, and 
a lower limit to the age of WD in M4 of $\sim$9~Gyr has been
derived from a comparison with theoretical
WD cooling curves (Richer et al 1997). Future access to
cooler and fainter objects will better constrain $\rm T_U$.
\end{itemize}
According to Sandage \& Tammann (1997) and Saha et al (1999) 
the Hubble constant $H_0$ should be in the range
$\rm 55\pm5 km.s^{-1}.Mpc^{-1}$, which implies
$T_U=\frac{2}{3}H_0^{-1}\approx11-13.5$ Gyr,
indicating that no strong discrepancy with the age of the oldest known stars remains.

\subsection{Variable Stars}
\label{puls}

I shall not discuss the revisions of the distance scale
based on pulsating stars (RR Lyrae, Cepheids, Miras,
high-amplitude $\delta$ Scuti stars) because this topic
has been extensively reviewed by Caputo (1998) and Reid (1999).

Both new insight as well as new questions about the
physics governing pulsating stars have been generated
from the combination of Hipparcos distances with asteroseismic data.
When the magnitude of a star is modified and its error box
reduced, the mass and evolutionary stage attributed to the star
may be modified. For variable stars, a different evolutionary
stage may give a drastically different eigenmode spectrum, and in
turn may change the mode identification and asteroseismic analysis
(see Liu et al 1997). H\o g \& Petersen (1997) showed that for
the two double-mode, high-amplitude $\delta$ Scuti variables 
SX Phe and AI Vel, the masses derived on one hand from stellar 
envelope models and pulsation theory, and on the other hand 
from the position in the H-R diagram through stellar evolution 
models are in nice agreement if the Hipparcos parallaxes (accurate to
5-6\%) are used. Further implications of Hipparcos distances on the
understanding of $\delta$ Scuti stars, $\lambda$ Bootis, and
rapidly oscillating Ap stars have been discussed in several papers
(see for instance Audard et al 1998, Viskum et al 1998, 
Paunzen et al 1998, Matthews et al 1999),
whereas the physical processes relevant to the Asymptotic Giant Branch
and pulsation modeling of Miras and Long Period Variables
were examined by Barth\`es (1998, see also references therein).

\subsection{White Dwarfs}
\label{WD}

The white dwarf (WD) mass-radius (M-R) relation was first derived
by Chandrasekhar (1931) from the theory of stars supported by the
fully degenerate electron gas pressure. It has
been refined by Hamada \& Salpeter (1961), who calculated
zero-temperature (fully degenerate) WD models of different chemical composition (He,
C, Mg, Si, S, and Fe) and by Wood (1995), who calculated WD models
with carbon cores and different configurations of hydrogen and/or helium layers
and followed the thermal evolution of WD as they cool.
Although theoretical support is strong, it has long been difficult to
confirm the relation empirically because of (1) the very few
available WD with measures of masses and
radii, (2) the size of the error bars and (3) the intrinsic mass
distribution of the WD, which concentrates them in only a small interval
around 0.6 \Msun~ (Schmidt 1996).

The M-R relation, assuming that WD have a carbon core, is a basic
underlying assumption in most studies of WD properties.
It serves to determine the mass of WD, and in turn their
mass distribution and luminosity function.
It is important because WD feature in
many astrophysical applications such as the calibration of
distances to globular clusters (Renzini et al 1996)
or the estimate of the age of Galactic disk and halo by means of WD
cooling sequences (see Winget et al 1987, d'Antona \& Mazzitelli 1990). 
The more precisely the M-R relation is defined
by observations, the better the tests of theoretical models of WD interiors
that can be undertaken. These include tests of the inner chemical composition
of WD, thickness of the hydrogen envelope of DA WD, or the
characterization of their strong inner magnetic fields.

Depending on the white dwarf considered, the empirical
M-R relation can be obtained by different means:

\begin{enumerate}
\item {\sl Surface brightness method.} If \Teff~ and \logg~ are
determined (generally from spectroscopy), then
model atmospheres allow calculation of the energy flux at the surface
of the star, which when compared to the flux on
Earth yields the angular diameter $\phi$ (see Schmidt 1996).
The radius R is obtained from the parallax and $\phi$, M is
deduced from R and \logg. This method requires high-resolution
spectra and largely depends on model atmospheres.

\item {\sl Gravitational redshift.} The strong gravitational
field at the surface of a WD causes a redshift of the spectral
lines, the size of which depends on the
gravitational velocity $v_{\rm grs}= \frac{G M}{R c}$
($c$ is the speed of light). If $v_{\rm grs}$ can be measured and
the gravity is known, then M and R can easily be obtained
independently of the parallax. $v_{\rm grs}$ can only be measured in WD
members of binary systems, common proper motion pairs (CPM), or
clusters because the radial velocity is required to
distinguish the gravitational redshift from
the line shift due to Doppler effect. Also, very high-resolution
spectra are needed.

\item {\sl WD in visual binary systems.} Masses may be derived directly
from the orbital parameters through the Kepler's third law,
provided the parallax is known. Radii are derived from the knowledge
of \Teff~ and distances.
\end{enumerate}

More than 15 years ago, when the Hipparcos project began,
uncertainties on WD ground-based parallaxes were at least 10 mas.
During the last 10 years, due to great instrumental
progress, parallax determinations were improved by
a factor of about 2, and more accurate atmospheric parameters \Teff,
\g\ and $v_{\rm grs}$ were obtained. In the meantime, Hipparcos observed
22 white dwarfs (11 field WD,  4 WD in visual binaries, and 7 in
CPM systems) among which 17 are of spectral type DA.
Although they are close to the faint magnitude limit of Hipparcos, the
mean accuracy on their parallaxes is $\sigma_\pi\simeq$3.6 mas
(Vauclair et al 1997).

Vauclair et al (1997) and Provencal et al (1998) studied the whole sample
of WD observed by Hipparcos. The M-R relation is
narrower and most points are within $1\ \sigma$ of Wood's (1995) evolutionary models of
WD with carbon cores and hydrogen surface layers.
The theoretical shape is still difficult to confirm
because of the lack of objects in the regions of either
high or low mass. Furthermore, the error bars are still too large to
distinguish fine features of the theoretical models, such as
between evolutionary and zero-temperature sequences or
thick and thin hydrogen envelopes (Vauclair et al 1997), except
for some particular stars (Shipman et al 1997, Provencal et al 1998).
Other effects such as alterations due to strong internal
magnetic fields are not yet testable (Suh \& Mathews 2000).

{\sl WD in binary systems.} Prior to Hipparcos,
Sirius B was the only star roughly located on the expected theoretical
M-R relations; the others (Procyon B, 40 Eridani B and Stein 2051) were
at least $1.5\sigma$ below the theoretical position
(see Figure 1 in Provencal et al 1998). After Hipparcos, as
shown by Provencal et al, the error on the radius is dominated
by errors on flux and \Teff. On the other hand, the parallax error
still dominates the error on mass, except for Procyon where the
error on the component separation plays a major role.

\begin{itemize}

\item Sirius B is more precisely located on a Wood's (1995) M-R relation
for a DA white dwarf of the observed \Teff~ with a thick H layer and
carbon core (Holberg et al 1998). Also compatible with Wood's
thick H layer models is the position of V471 Tau, a member of an
eclipsing binary system for which the Hipparcos parallax 
supports the view that it is a member
of the Hyades (Werner \& Rauch 1997, Barstow et al 1997).
The mass of 40 Eri B increased by 14\%. The star is now back
on Hamada \& Salpeter's (1961) M-R relation for carbon cores,
making it compatible with single star evolution
(Figure 1 by Shipman et al 1997) and it does not appear to have a
thick H layer.
Sirius B (the most massive known WD) and 40 Eri B (one of the
less massive) nicely anchor the high-mass and low-mass limits of the
M-R relation.

\item The case of Procyon B remains puzzling. The position is
not compatible with models with carbon cores, and would be better
accounted for with iron or iron-rich core models (Provencal et al 1997).
Provencal et al (1998) also examined seven white dwarf members of CPM pairs
(more distant and fainter than WD in visual binaries) with
Hipparcos distances and gravitational redshift measurements. They
showed that two of them also lie on theoretical M-R relations
corresponding to iron cores. This is not predicted by current
stellar evolution theory, and further work is required to clarify
this problem.

\end{itemize}

In conclusion, better distances from Hipparcos and high-resolution
spectroscopy has allowed better assessment of the theoretical M-R relation
for white dwarfs, and has shown evidence for difficulties for a few
objects that do not appear to have carbon cores. Future
progress will come from further parallax improvements and
from better \Teff, $v_{\rm grs}$, magnitudes, and orbital parameters
for visual binaries. A better understanding of the atmospheres is
required: convection plays a role in the cooler WD, additional
pressure effects due to undetectable helium affect the gravity
determination, and incorrect H layer thickness estimates change the mass
attributed to WD. Further information coming from
asteroseismology or spectroscopy would help.

\section{FUTURE INVESTIGATIONS}
\label{future}

Hipparcos has greatly enlarged the available stellar samples
with accurate and homogeneous astrometric and photometric data.
To fully exploit this new information, many studies have been 
undertaken (several hundred papers devoted to stellar studies
based on or mentioning
Hipparcos data can be found in the literature), therefore
the present review could not be fully exhaustive.

The Hipparcos mission succeeded in clarifying our knowledge
of nearby objects, and allowed first promising studies of rarer or
farther objects. After Hipparcos, the theory
of stellar structure and evolution is further anchored, and
some of its physical aspects have been better characterized.
For instance, new indications that the evolution of
low-mass stars is significantly modified by microscopic diffusion
have been provided by fine studies of the H-R diagram,
and consequences for age estimates or surface abundance
alterations have been further investigated. 
On the other hand, Hipparcos left us with intriguing results that
raised new questions. For example, the unexpected position of
the white dwarf Procyon B on a theoretical mass-radius relation
corresponding to iron cores is still not understood. 

Today, uncertainties on distances of nearby stars have been reduced
significantly such that other error sources emerge to dominate,
hindering further progress in
the fine characterization of stellar structure.
Progress on atmosphere modeling is worth being pursued, for it has
implications for observational parameters (effective temperatures,
gravities, abundances, bolometric corrections), theoretical
models (outer boundary conditions), and color calibrations.
A thorough theoretical description of transport processes (convection,
diffusion) and related effects (rotation, magnetic fields) is
needed to improve stellar models, as well as further
improvements or refinements in microscopic physics (low-temperature
opacities, nuclear reaction rates in advanced evolutionary stages). 

What is now needed from the observational side is (1)
enlarged samples of rare objects (distant objects,
faint objects or objects undergoing rapid evolutionary phases),
(2) an increased number of more ``common'' objects 
with extremely accurate data (including masses), and (3) a census
over all stellar populations.

These goals should be (at least partially) achieved by future
astrometric missions. The NASA Space Interferometry Mission (SIM),
scheduled for launch in 2006, will have the capability to measure
parallaxes to 4 $\mu$arcsecond and proper motions to 1-2
$\mu$arcsecond per year down to the $20^{\rm th}$
magnitude which represents a gain of three orders of magnitude with respect to
Hipparcos (Peterson \& Shao 1997).
The ESA candidate mission GAIA is dedicated to the observation of
about one billion objects down to
V$\simeq$20 mag (and typical $\sigma_\pi \sim$10
$\mu$arcsecond at V=15 mag). GAIA will also provide multi-color, multi-epoch
photometry for each object, and will give access to stars of various
distant regions of the Galaxy (halo, bulge, thin and thick disk, spiral
arms). It is aimed to be launched in 2009 (Perryman et al 1997b). 

Asteroseismology has already proved to be a unique tool
to probe stellar interiors. Space experiments are under study.
The first step will be the COROT mission (aimed to be launched in 2003),
designed to detect and characterize oscillation modes
in a few hundred stars, including solar-type
stars and $\delta$ Scuti stars (see Baglin 1998).
  
\section{Acknowledgments}

I enjoyed very much working with Ana G\'omez, Roger and Giusa Cayrel,
Jo\~ao Fernandes, Michael Perryman, No\"el Robichon, Annie
Baglin, Jean-Claude Mermilliod, Marie-No\"el Perrin,
Fr\'ed\'eric Arenou, Catherine Turon, and Corinne Charbonnel
on the various subjects presented here. I particularly thank 
Roger Cayrel, Annie Baglin, Michael Perryman,
and Catherine Turon for a careful reading of the original manuscript. I 
am also very grateful to Allan Sandage for his
remarks and suggestions which helped improving the paper.
I am especially grateful to Fr\'ed\'eric Th\'evenin, Hans-G\"unter
Ludwig, Misha Haywood, Don VandenBerg, Marie-Jo Goupil, Pierre
Morel, Jordi Carme and Franck Thibault for fruitful discussions and advice.
The University of Rennes 1 is acknowledged for working facilities.
This review has made use of NASA's Astrophysics Data System
Abstract Service mirrored at CDS (Centre des Donn\'ees
Stellaires, Strasbourg, France).

\end{document}